\documentclass[%
 reprint,
 amsmath,amssymb,
 aps,
]{revtex4-2}

\usepackage{graphicx}
\usepackage{dcolumn}
\usepackage{bm}

\usepackage{color}

\begin{document}

\preprint{APS/123-QED}

\title{Partially bonded crystals:\\ a pathway to porosity and polymorphism}

\author{Carina Karner}
 \email{Carina.Karner@tuwien.ac.at}
\affiliation{Institut f\"{u}r Theoretische Physik, TU Wien, Wiedner Hauptstraße 8-10, A-1040 Wien, Austria}
\author{Emanuela Bianchi}%
 \email{Emanuela.Bianchi@tuwien.ac.at}
 \affiliation{Institut f\"{u}r Theoretische Physik, TU Wien, Wiedner Hauptstraße 8-10, A-1040 Wien, Austria and CNR-ISC, Uos Sapienza, Piazzale A. Moro 2, 00185 Roma, Italy}

\date{\today}

\begin{abstract}
In recent years, experimental and theoretical investigations have shown that anisotropic colloids can self-organise into ordered porous monolayers, where the interplay of localised bonding sites, so called patches, with the particle's shape is responsible for driving the systems away from close-packing and towards porosity. Until now it has been assumed that patchy particles have to be fully bonded with their neighbouring particles for crystals to form, and that, if full bonding cannot be achieved due to the choice of patch placement, disordered assemblies will form instead. In contrast, we show that by deliberately displacing the patches such that full bonding is disfavored, a different route to porous crystalline monolayers emerges, where geometric frustration and partial bonding are pivotal in the structure formation process. The resulting dangling bonds lead to the emergence of effectively chiral units which then act as building blocks for energetically equivalent crystal polymorphs. 
\begin{description}
\item[Usage]
Electronic Supplementary Information (ESI) available. 
\end{description}
\end{abstract}

\maketitle


\section{\label{sec:level1} Introduction}

With recent advancement in colloidal synthesis techniques it has become possible to fabricate colloidal particles with precise control over their size, shape and interaction profiles  \cite{pawar2010fabrication, bianchi2011patchy, Sacanna2011, walther2013janus, Boles2016, Small_2018}. 
Via self-assembly, these nano- to micron-sized building blocks can be used for bottom-up materials design, where properties of the self-assembled material can be engineered by fine-tuning the particle characteristics.
A particularly relevant subclass of colloidal particles in this context are patchy particles, where only specific sections of the particle's surface are functionalised~\cite{pawar2010fabrication, bianchi2011patchy, Ravaine2013, sacanna2013shaping, ravaine2017synthesis, wang2012colloids, he2020colloidal, chen2011directed, diaz2020photo, oh2019colloidal, zhang2017janus, liu2016self,klinkova2013colloidal, iubini2020aging, sciortino2007self, doppelbauer2010self, doye2007controlling, reinhardt2013computing,liu2019rational}. This selective functionalization allows the particles to bond exclusively at these "patchy" regions, making it possible to exploit the intuitive connection between the patch bonding pattern and the target crystal symmetry.

In experiments, it has been shown that the assembly of spherical units is largely controlled by the location and size of the patches, as demonstrated by polystyrene particles with hydrophobic gold patches on both poles \cite{chen2011directed}. These so-called triblock particles assemble into a porous Kagome lattice as long as the size of the polar patches allows allocating two bonds per patch. When this condition is satisfied, bonding is maximized in a triangular local geometry, leading to the extended Kagome monolayer.
Similar patchy design principles have been successfully exploited to  assemble the highly desired colloidal diamond from particles with tetragonal patch symmetry~\cite{he2020colloidal, morphew2018programming} . 
For small patches in the one bond per patch limit,  the intuitive relationship between patch number, size and placement on the particle surface becomes even more apparent.  As small patches on spherical units are still hard to achieve in experiments, one has to rely also on simulations. These show that one patch particles typically assemble dimers~\cite{oh2019colloidal,klinkova2013colloidal}, two patch systems assemble finite clusters or chains~\cite{iubini2020aging, sciortino2007self, wang2012colloids}, while extended structures have only been observed starting from three patches~\cite{doppelbauer2010self}, where the symmetry of the patch arrangement directly impacts the symmetry of the assembly. Examples are two-dimensional assemblies where spherical particles with four patches placed along the sphere perimeter at $90$\textdegree intervals assemble into a square lattice or where particles with six evenly distributed patches yield hexagonal crystals~\cite{doye2007controlling, doppelbauer2010self}.   
With the same rationale even dodecagonal quasicrystals can be assembled by placing five patches equally spaced along the particle perimeter as this creates a local neighbourhood with five-fold symmetry required for the growth of quasicrystals~\cite{reinhardt2013computing,liu2019rational}. 

Besides number and size of patches, changing the particle shape from spherical to aspherical adds another dimension for parameter exploration. When the shape of the particles is anisotropic, some mutual particle orientations might be favoured with respect to others on a purely entropic basis. The interplay between shape and patchiness creates a delicate balance between entropic and enthalpic factors opening up more possibilities for intricate particles arrangements. Examples of assemblies resulting from such a delicate balance are non-spherical colloidal platelets with functionalized edges or localized ligands, such as nanocrystals, DNA origami and polymer-based particles  \cite{Ye2013, Millan2014, Lee_2016, Whitelam2012, Karner-Nanolett2019, pakalidou2020engineering, Cai2018hierarchical, kocabey2015membrane, journot2019modifying, Qian_2018}

These lego-like design principles are also being exploited for inverse design, where given the target structure, the building block shape, and the patch size and position are determined by using Machine Learning algorithms~\cite{ma2019inverse, long2018rational, chen2018inverse, lieu2022inverse, rivera2023inverse, truong2024prediction}.

Ultimately, be it by intuiting how patchy particles bond locally or by employing  optimizing strategies, the underlying assumption remains -- by design -- that the desired crystalline or quasi-crystalline target structures result from all patches being bonded to at least one other patch. 
This raises a central question: what happens to the self-assembly of patchy particles if they are prevented from fully bonding due to conflicting interactions? 

Here, we focus on a class of patchy platelets with the shape of regular rhombi and four localized bonding sites (one per edge) of two types: while patches of the same type attract each other, patches of opposite types repel each other; where bond selectivity is tuned by considering different arrangements of the two patch types along the platelets' perimeter. 
Extensive investigations have shown that these patchy rhombi tend to assemble in fully bonded, close-packed tilings as long as their patches are placed in the center of their edges, irrespective of their bond selectivity pattern~\cite{Karner-Nanolett2019,karner2020patchiness,karner2024anisotropic}. In contrast, when the patches are placed off-center, a complex interplay sets in, where the pairwise alignment resulting from the particle shape opposes the enthalpic drive to maximize the bonding, often leading to off-edge bonding between the particles~\cite{Karner-Nanolett2019}. The formation of off-edge bonds opens up the possibility for porous assemblies that can be either fully~\cite{Karner-Nanolett2019,karner2020patchiness} or partially~\cite{karner2024anisotropic} bonded, 
depending on the interplay between patch position and bond selectivity. While fully bonded open assemblies usually result in crystalline porous monolayers, where porosity is controlled by the degree of off-edge bonding between the particles~\cite{Karner-Nanolett2019,karner2020patchiness}, partially bonded assemblies lead to the formation of extensive disordered networks~\cite{karner2024anisotropic}. Even though the emergence of these disordered networks seems to be a robust scenario in all patchy rhombi systems where geometric bond frustration is present~\cite{Karner-Nanolett2019,karner2024anisotropic}, the emergence of non-fully bonded crystals cannot be in principle ruled out.  

\section{Results}
\begin{figure}
\includegraphics[width=\columnwidth]{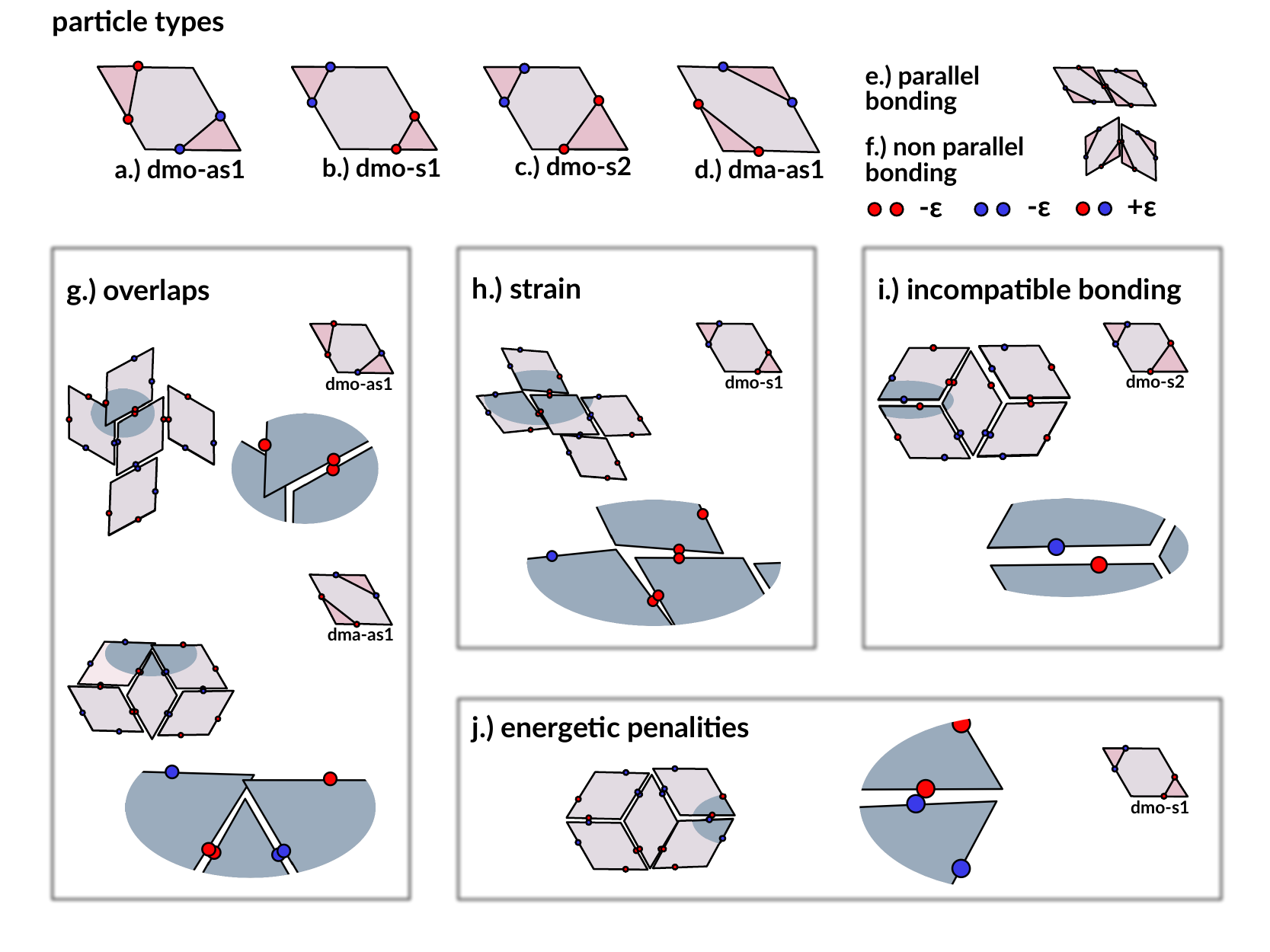}
\caption{{\bf Particle types, pair bonding motifs and full bonding failures.} All studied particle types and example bonding motifs where full bonding either fails or is disfavored. To facilitate the distinction between the particle types, we highlighted the triangle area spanned by  patches of the same type with their enclosed edge in a darker shade. The particle types are (a) dmo-as1 (b) dmo-s1,  (c) dmo-s2, (d) dma-as1. In (e) and (f) we depict the two modes of bonding available for rhombus particles: parallel (e) and non-parallel (f).  In (g) we show motifs with overlaps for dmo-as1 and dma-as1; (h) depicts a dmo-s1 motif where full bonding is possible only under strain; (i) shows an instance for incompatible bonding for dmo-s2, where relative patch placement prevents full bonding; in (j) we depict how the bonding motif can force patches of different type to face each other, resulting in an energetic penalty, as exemplified by dmo-s1.}
\label{fig:strained_particle_types}
\end{figure}

Leveraging on the described framework and on our experience, we consider a subset of four arbitrary, but carefully selected, particle types, namely dmo-as1, dmo-s1, dmo-s2 and dma-as1, as shown in Figure~\ref{fig:strained_particle_types}a-d, where the particle naming scheme leans on previous nomenclature from Ref.~\cite{Karner-Nanolett2019} and is specified in the Methods section, where details of the model features and parameters are also reported. Despite the displacement of the patches with respect to their central position being relatively small (as it amounts to $10\%$ of the length of the rhombi edge), the patch arrangement and bond selectivity, together, strongly disfavour or even prevent the formation of fully bonded assemblies. In particular, the investigated systems have been selected to introduce geometric bond frustration via at least one of the full bonding failures depicted in Figure~\ref{fig:strained_particle_types}g-j (for a more exhaustive  set of failing bonding motifs see Ref.\cite{karner2024anisotropic}): in the selected systems, the four patches of a particle cannot simultaneously bond to the neighbouring particles without incurring either overlaps, incompatible bonding (due to steric hindrance), energetic penalties (due to patches repelling each other), or at least a certain degree of strain induced by patches bonding almost out of interaction range. 
In order to robustly assess the effect of geometric bond frustration on self-assembly, we conducted large scale real-space Monte Carlo simulations performing a substantial total of $24576$ separate simulations -- considering all $16$ parallel runs for each of the $384$ state points across all $4$ particle topologies --  lasting from $3$ to $4\times 10^7$ Monte Carlo sweeps (see the Methods section for more details).  

\begin{figure}
\includegraphics[width=\columnwidth]{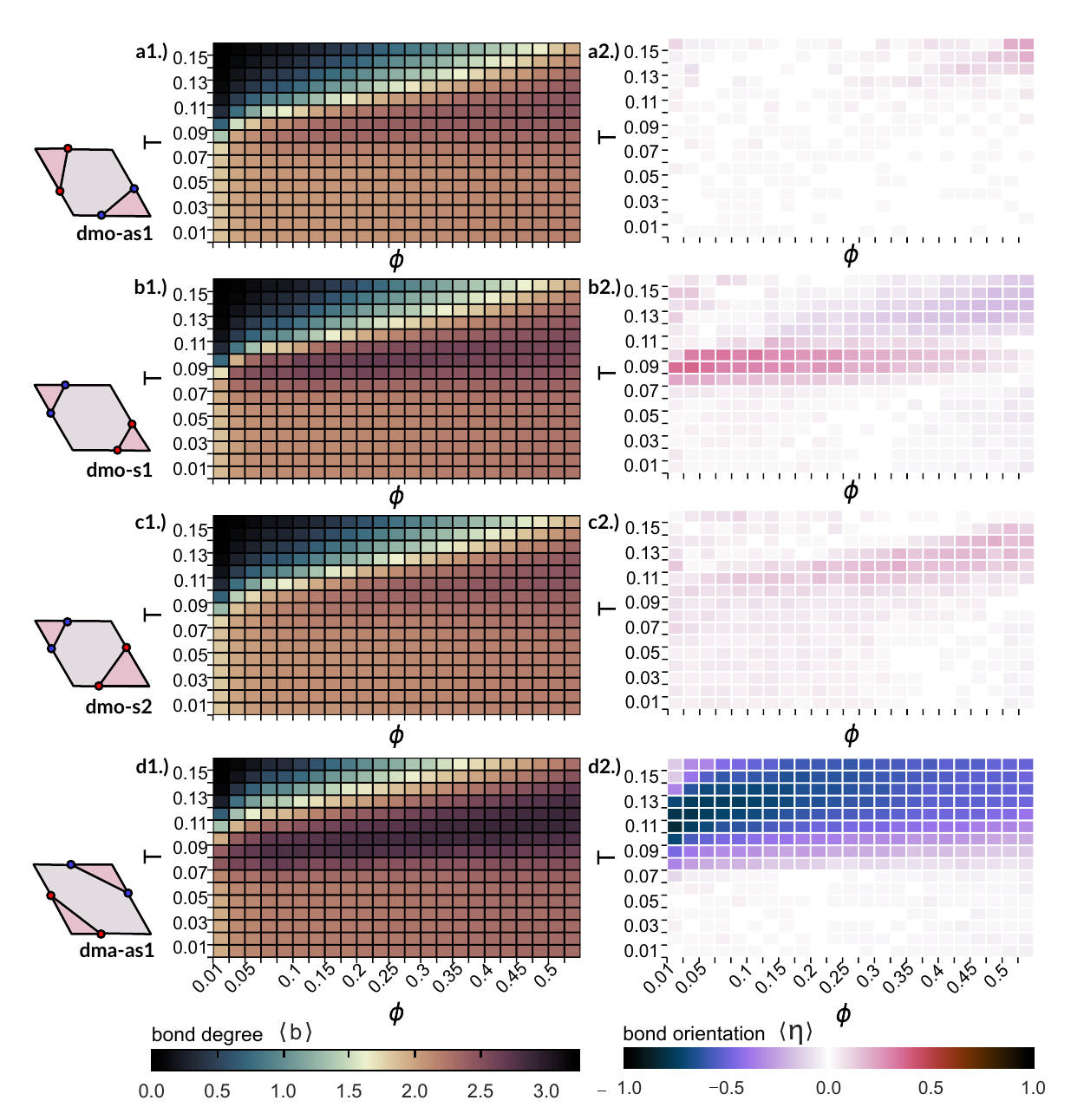}
\caption{{\bf Average number of bonded neighbours (left) and average  bond orientation (right).} Heat maps for the average number of bonded neighbours, $\langle b \rangle$, and the average bond orientation, $\langle \eta \rangle$, across all investigated state points (with $\phi=0.05-0.525$ and $T=0.01-0.16$) for all studied particle types. At each data point averages are obtained according to the standard statistics procedure (see Methods section). (a1-d1)  Heat maps for $\langle b \rangle$ for dmo-as1 (a1), dmo-s1 (b1), dmo-s2 (c1) and dma-as1 (d1), where the minimum is zero bonds (dark blue) and the maximum is four bonds (dark brown).  (a2-d2) Heat maps for $\langle \eta \rangle$ for dmo-as1 (a2), dmo-s1 (b2), dmo-s2 (c2) and dma-as1 (d2), where $\langle \eta \rangle = -1/+1 $ denotes the extremes where all bonds are non-parallel (dark blue)/ parallel (dark red) and $\langle \eta \rangle = 0$ indicates state points with mixed bonding on average (white).}
\label{fig:average_bonding}
\end{figure}

{\bf Bonding properties at the particle level.}
A first insight into the assembly scenarios of the selected systems is provided by the average number of bonding neighbours per particle, $\langle b \rangle$, along the chosen grid of state points.  
The results for all patch topologies are summarized in Figure~\ref{fig:average_bonding}a1-d1 via heat maps in the temperature ($T$) $versus$ packing-fraction ($\phi$) plane. 
 We observe that, for the highest $T$,  at all but the highest $\phi$-values, particles bond to $0-1.5$ neighbours on average, which indicates a liquid state where particles are either not bonded or bonded within small clusters. At $T<0.09$,  the average bonding is  $2.1-2.2$ across all packing fractions, which confirms the network character of the assemblies: here, the majority of particles forms linear chains bonding two patches,  that are interconnected with a few branching elements characterized by three bonds. 
The behaviour of $\langle b \rangle$ in the highlighted high and low $T$-regimes fully supports the picture provided by our previous investigations of these four patch topologies, which  revealed a liquid state at high $T$ and -- separated by a percolation line --  a disordered particle network at low $T$, which we fully characterized by bonding motifs, porosity and percolation loci~\cite{karner2024anisotropic}.
However, the heat maps of $\langle b \rangle$ across temperature and density show very clearly that, for all four patch topologies, there is a region between the liquid and the network state, where the average bonding exceeds the network bonding of 2.2 and reaches values between 2.3 and 3. 
While these regions of high bonding do significantly differ in extent in the $T-\phi$ plane as well as in the maximum average bonding across the four patch topologies, we observe that, generally, they do occur at intermediate temperatures and at intermediate to high densities. 
These domains of higher bonding suggest that the assembled structures in this region may be qualitatively different from the liquid  at high and the networks at low temperature. Nonetheless, on the basis of the parameter $\langle b \rangle$ only, a characterization of the assemblies cannot be performed, since this higher bonding might equivalently point to ordered crystals, finite clusters or even to just more tightly bonded disordered networks.

In an attempt to clarify further whether these assemblies are ordered or disordered, we evaluate the orientational order of the aggregates by estimating the average orientation of bonded particles $\langle \eta \rangle$: as two patchy rhombi can bond either parallel or non-parallel (see Figure~\ref{fig:strained_particle_types}e and f for illustration), the bond orientational parameter can be calculated using the percentage of either bonding orientation. In this case,  we define $\langle \eta \rangle = 1-2P_{np}$, where $P_{np}$ is the fraction of non-parallel bonds, which means that $\langle \eta \rangle$ takes values between -1 and 1, where a value of -1  indicates that the assembly is bonded entirely in a non-parallel fashion, while a value of 1 indicates a completely parallel assembly and a value of 0 indicates a mixed assembly with an equal amount of parallel and non-parallel bonds. 
In Figure~\ref{fig:average_bonding}a2-d2, where we show the full heat maps  for $\langle \eta \rangle$, shades of blue/red represent assemblies with more non-parallel/parallel bonding, while assemblies with an equal mix of parallel and non-parallel bonds are indicated by the white color.  

While, across all dmo-systems, the networks and fluid states predominantly show mixed bonding, in the region of high bonding dmo-s1 and dmo-s2 exhibit a clear tendency towards parallel or non-parallel bonding (Figures~\ref{fig:average_bonding} b2 and c2 ), while dmo-as1  remains of mixed bonding also in the high bonding zone (Figure~\ref{fig:average_bonding}a2).  
Notably, in the dma-as1 system, a pronounced preference for non-parallel bonding is evident even in the fluid state, persisting through the region of high bonding and changing to mixed bonding only for the disordered network state  (Figure~\ref{fig:average_bonding}d2). 
It is worth noting, that, while the observed bond orientational order for dmo-s1, dmo-s2 and dmo-as1 may stem from crystallites with orientational and positional order, it could be also due to small clusters with repeating bonding motifs, either within a fluid state or bonded within an otherwise disordered network. 
Vice versa, the predominantly mixed bonding for dmo-as1 may indicate an orientationally disordered assembly, but could also signify  orientational order with a more complex bonding pattern. 
We conclude that, while the heat maps for $\langle b \rangle$ and $\langle \eta \rangle$ suggest the presence of assembly scenarios distinct from the previously studied liquid and network states, only a broad visual analysis followed by a rigorous crystal structure detection would bring clarity whether these assemblies are ordered or disordered. 

\begin{figure*}
\includegraphics[width=\textwidth]{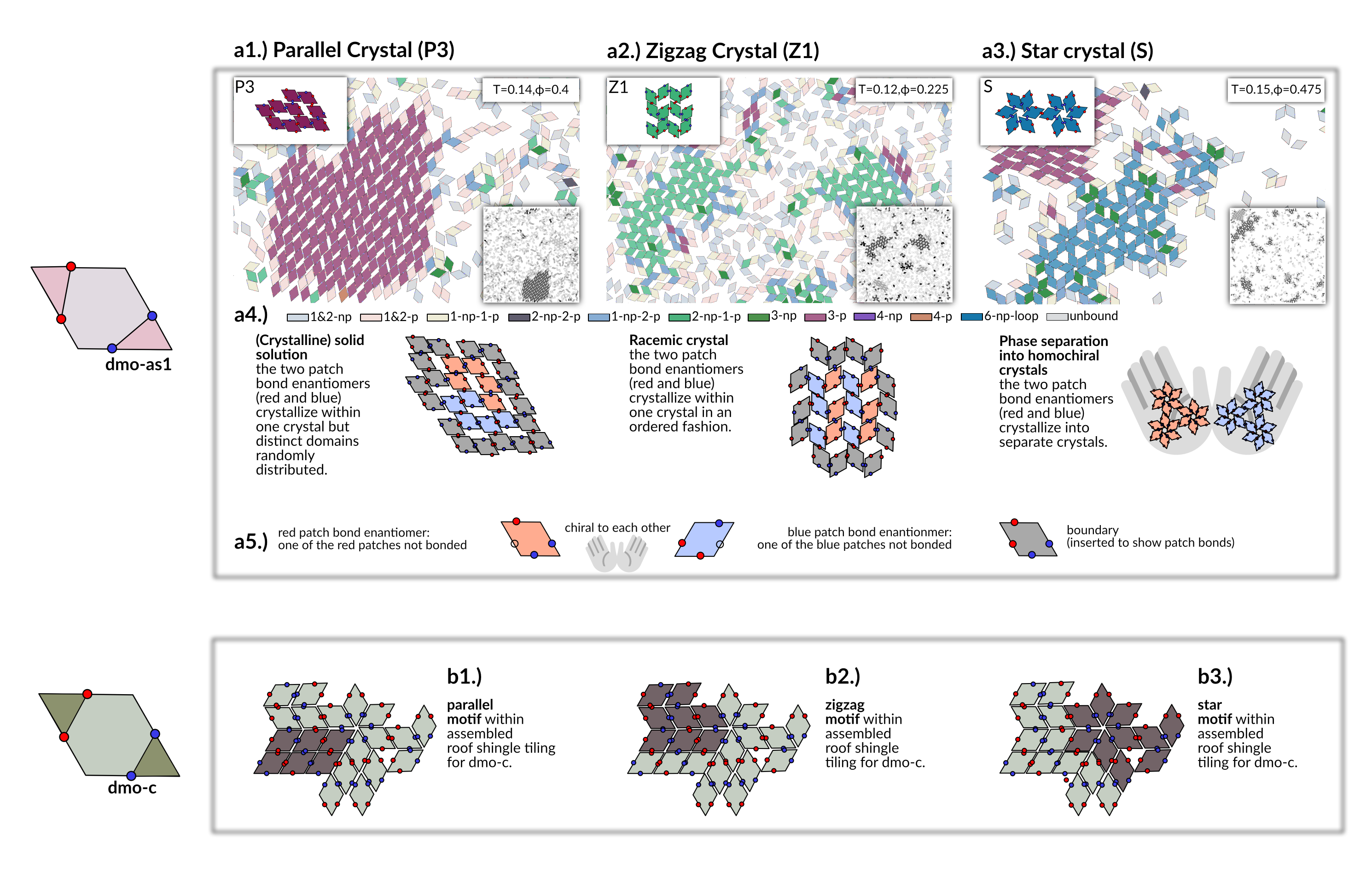}
\caption{{\bf Crystal polymorphs (top) emerging from the geometric frustration of a close-packed tiling (bottom).} Snapshots and sketches of emerging crystal polymorphs for dmo-as1 (a-panels) in comparison to bonding motifs in the roof-shingle tiling observed for dmo-c (b-panels), where all patches are in the center of their respective edge~\cite{karner2024anisotropic, Karner-Nanolett2019}. Particles within the simulation snapshots are colored according to the number and orientations (parallel, p, or non-parallel, np) of their bonds. The sketches below the snapshots highlight that all depicted crystals are formed by particles with three out of four patches bonded: either two blue and one red bond (red particles) or two red and one blue bond (blue particles); red and blue particles are chiral to each other (see legend in a5).  (a1) Parallel crystal (P3), where each particle has three p-bonds, snapshot taken at $\phi=0.4$, $T=0.14$.  The sketch below the snapshot illustrates that P3 is a solid solution of red and blue particles. (a2) Zigzag crystal (Z1), where each particle has two np- and one p-bond, snapshot taken at $\phi=0.225$, $T=0.12$. The sketch below the snapshot shows that Z1 is a racemic crystal with an equal amount of red and blue particles. (a3) Star crystal (S), where each particle has two np- and one p-bond, where the np-bonds are arranged in a loop of six np-particles (star), snapshot taken at $\phi=0.475$, $T=0.15$. The sketch below the snapshot shows that S appears in two homochiral forms - one form containing only the red particles and one only containing the blue particles.
(a4) Legend for coloring of particles in the snapshots. 
The naming  generally follows the scheme a-np-b-p, where a is the number of np-bonds and b is the number of p-bonds. Exceptions are 1$\&$2-np/p which refers to particles with only one or two np-/p-bonds and 6-np-loops which refers to particles within a closed loop of six np-particles. 
(b1) Sketch highlighting the fully bonding parallel motif within the observed roof-shingle tiling for dmo-c. (b2)  Sketch highlighting the zigzag motif with the observed roof-shingle tiling for dmo-c. (b3) Sketch highlighting the star motif within the observed roof-shingle tiling dmo-c.}
\label{fig:crystals_dmo_as1}
\end{figure*}

\begin{figure*}
\includegraphics[width=\textwidth]{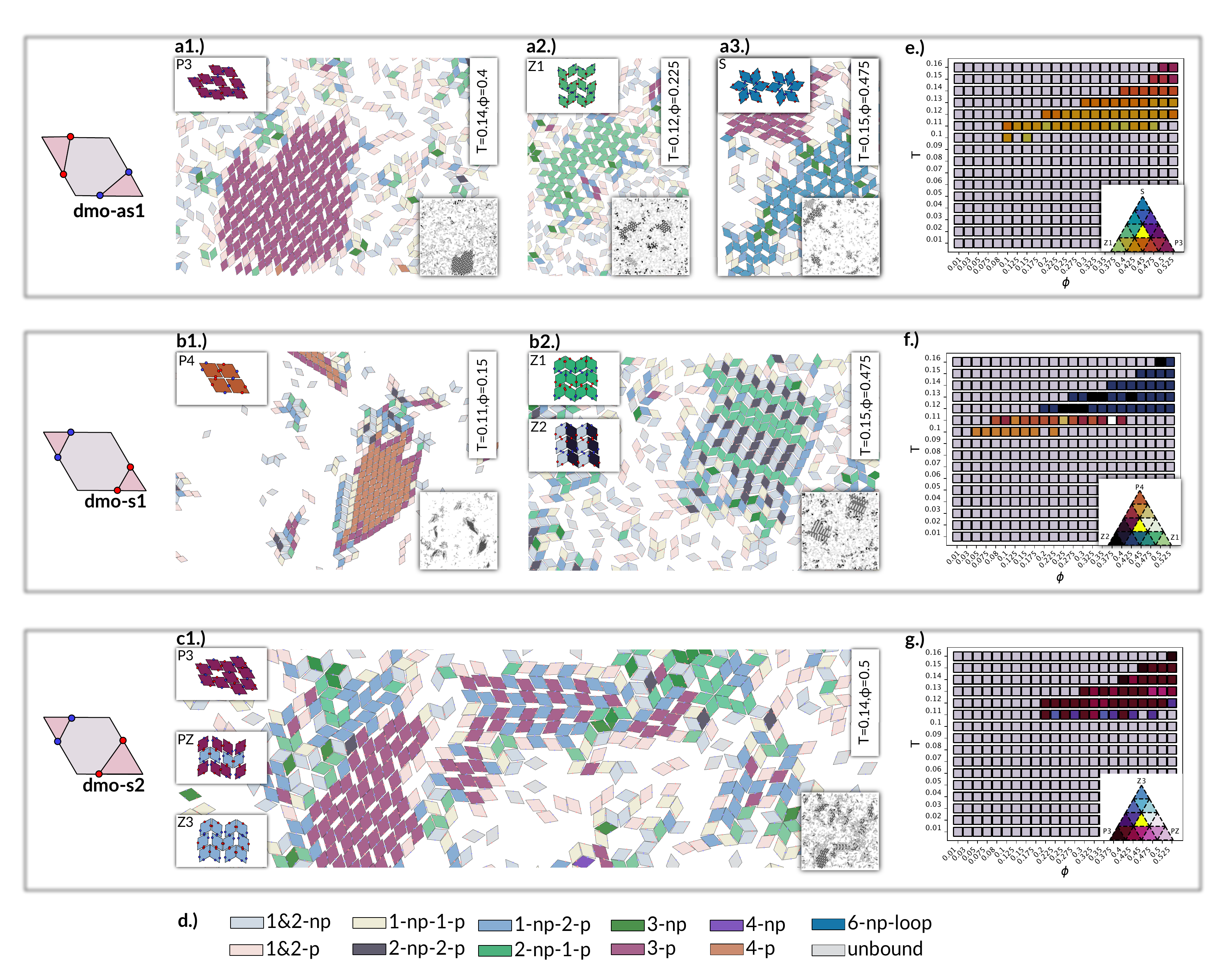}
\caption{{\bf Overview of the observed crystal polymorphs and their relative abundance.} (a-c) Snapshots and sketches of emerging crystal structures for dmo-as1, dmo-s1 and dmo-s2 as well as (e-g) heat maps quantifying their relative abundance across different state points $(\phi,T)$. Particles in snapshots are colored according to the number and orientation of bonded neighbours (see legend in (d) with the labeling scheme reported in Figure~\ref{fig:crystals_dmo_as1}), the corresponding values of $\phi$ and $T$ of each snapshot are reported by the labels. The crystal lattices are sketched at the left corner of each snapshot, while at the bottom right corner the corresponding zoomed-out, gray-scale snapshot is reproduced. (a) dmo-as1: (a1) parallel crystal P3, (a2) zigzag crystal Z1, (a3) star crystal S. (b) dmo-s2: (b1) fully bonded parallel lattice P4, (b2) zigzag lattices Z1 and Z2 for dmo-s1. (c1) Parallel lattice (P3) and zigzag lattices PZ and Z3. (e-g) Heat maps showing the relative abundance of the three most dominant crystal structures for each particle type across all state points ($\phi=0.01-0.525$, $T=0.01-0.16$).  Relative percentages were obtained by the crystal structure detection algorithm and averaged according to the standard statistics procedure (see Methods section). Color mapping is conducted via a the ternary color maps on the lower right corner, where grey indicates state points where percentage of crystalline particles of any kind is below 0.05. For each data point the averages were obtained according to the standard statistics procedure (see Methods section). 
(e) Relative abundance heat map for most dominant crystals (e) P3,Z1 and S for dmo-as1, (f) P4, Z2 and Z1 for dmo-s1, (g) P3, PZ and Z3 for dmo-s2.}
\label{fig:crystals_all}
\end{figure*}

{\bf Frustration-induced polymorphism.} We therefore visually inspect the simulation snapshots from all available state points for all four systems and  conclude that for three of them -- the dmo-systems  -- we do observe clusters of orientational and positional order -- i.e crystallites --  in the state point regions of higher bonding.  We visually detect three crystal polymorphs per particle type for a total of nine different crystals and we report them all in Figures~\ref{fig:crystals_dmo_as1} and~\ref{fig:crystals_all},  where  Figure~\ref{fig:crystals_dmo_as1} discusses the three polymorphs of dmo-as1 in detail, and Figure~\ref{fig:crystals_all} shows an overview of all found crystal structures. 
The majority of these crystals is novel and, to the best of our knowledge, has never been reported before. Therefore, in the absence of an established local order parameter to characterise these novel crystal structures, we craft our own measure. Inspired by the heat maps in Figure~\ref{fig:average_bonding}, where both, $\langle b \rangle$ and $\langle \eta \rangle$ show a characteristic variability  in the intermediate temperature region of interest, we calculate the number of parallel (p) and non-parallel (np) bonds for each particle  and color the snapshots in Figures~\ref{fig:crystals_dmo_as1} and~\ref{fig:crystals_all} accordingly.  All occurring bonding patterns and their respective coloring can be found in the legends in the Figures~\ref{fig:crystals_dmo_as1}a4 and  ~\ref{fig:crystals_all}d.
The resulting colored snapshots show that the chosen order parameter already successfully distinguishes between ordered domains, where the total number of bonds per particle tends to be higher as indicated by a darker color, and disordered domains, where the number of bonds tends to be lower -- represented by lighter colors.  For this reason, the number of p- and np-bonds per particle will serve as the basis for the crystal structure detection, to which we come later when we quantify the crystallinity  of the assemblies and the relative abundance of the polymorphs for different state points. 

By analyzing the colored snapshots for dmo-as1 (see the example reported in Figure~\ref{fig:crystals_dmo_as1}) we identify a parallel lattice P3 (in burgundi), a zigzag lattice Z1 (in green) and a star lattice S (in blue). Despite being characterised by higher bonding, the emerging crystals for the dmo-as1 system are not fully bonded. In fact, due to occurring overlaps when attempting to bond all four patches they cannot at all form fully bonded assemblies. 
Instead,  all three occurring  polymorphs are what we refer to as partially bonded by design, where the crystal lattices are built up from only a subset of all available patch bonds, while the rest of the patches remains unbonded. In the case of dmo-as1 only three out of four patches per particle contribute to the crystal lattices. 
In Figure~\ref{fig:crystals_dmo_as1}a1 we observe a large cluster of P3 which is built up from particles connected by three parallel bonds (3-p). 
The absence of the fourth bonded neighbour creates space and uniform rhombic pores form within the P3 lattice. 
Similarly, the other two polymorphs of dmo-as1, the Z1 and the S lattices, also consist of three contributing bonds per particle,  and again, the lack of the fourth neighbour particle results in the formation of regular pores.  Although the bonding pattern of the Z1 and the S lattice is the same with two np-bonds and one p-bond, the lattice structures differ, with the Z1 lattice essentially consisting of np-connected zigzagging rows connected to each other with p-bonds opening up the space for rhombic pores, while the S lattice is built up by 6-particle np-loops -- stars -- connected to each other by p-bonds, giving rise to large triangular pores between the stars and smaller, hexagonal pores in the center of the stars. 

All the crystals emerging in the dmo-as1 system are partially bonded with only three out of four patches per particle forming a bond, where the patch that remains unbonded varies. In the sketches below the snapshots, we highlight this variation in bonding by coloring the particles: in red when two blue patches and one red patch are bonded, in blue when two red patches and one blue patch are bonded. The red and the blue particles are not identical but chiral to each other through their bonding pattern (see the legend of Figure~\ref{fig:crystals_dmo_as1}a5) and thus in the following  we refer to them as patch bond enantiomers. 

In the P3 lattice, both bond enantiomers coexist within the lattice and bond seamlessly to each other.  Due to bonding constraints, particles of one chirality are bonded to at least one neighbour of the same chirality, and thus form rows of homochiral domains, leading to two distinct pore orientations within the assembly.  
Within the context of organic chemistry, these connected but distinct domains of the red and blue enantiomers are known as (crystalline) solid solutions ~\cite{xiouras2018applications, sogutoglu2015viedma}.
In the Z1 lattice the situation is different, as the zigzagging rows are assembled from alternating red and blue patch bond enantiomers, and the rows are connected to each other alternately by red or blue particles. With an equal amount of red and blue enantiomers co-crystallizing, we classify this assembly as a racemic crystal~\cite{jacques1981enantiomers}.
On the contrary, S crystals consist solely of one type of patch bond enantiomer, with the lattice's overall chirality determined by the stars. Typically, stars form six-particle non-parallel loops where particles are bonded through only one type of patch. Consequently, if the stars are constructed from blue patches, the lattice comprises particles with two blue patches bonding within the stars, and one red patch bonding to connect the stars in a parallel fashion. This results in a crystal exclusively composed of the red patch bond enantiomer, where stars formed by red patches - meaning blue patch bond enantiomers - cannot attach. These blue and red S crystals are chiral to each other as well, and because they cannot bond to each other, we characterise them as phase-separated homochiral  (enantio-pure) crystals~\cite{lorenz2014processes}. 
In summary, within every polymorph of dmo-as1, partial bonding leads to chiral variations in bonding patterns.  Analogously to organic compounds, these chiral bonding patterns give rise to racemic assemblies, that -- within a full bonding scenario -- could only be realized with binary mixtures of chiral building blocks \cite{sogutoglu2015viedma,jacques1981enantiomers, lorenz2014processes}. 

Having examined the intricate bonding patterns arising within each polymorph, we now focus on how a single building block --  in this case the dmo-as1 - can give rise to three different crystalline polymorphs. We find that this problem is best investigated through a comparison with the bonding motifs observed in dmo systems where all patches are placed at the center of each edge (referred to as dmo-c in Figure~\ref{fig:crystals_dmo_as1}b1-b3). In this case, particles assemble into a close-packed lattice labeled as roof-shingle tiling~\cite{Karner-Nanolett2019}. Within such a lattice, we find full bonding analogs of each dmo-as1 polymorph bonding pattern, as illustrated by the sketches: the parallel motif (Figure~\ref{fig:crystals_dmo_as1}b1), the zigzag motif (Figure~\ref{fig:crystals_dmo_as1}b2) and the star motif (Figure~\ref{fig:crystals_dmo_as1}b3).
In  dmo-c, the central patch position results in on-edge bonding, aligning the particle edges completely. Consequently, all three bonding motifs are commensurable, leading to a disordered tiling with all three motifs present at once.  It is only by slightly shifting the patches off-center to the dmo-as1 patch topology that lattice selectivity arises where the different bonding motifs are not compatible anymore, leading to the formation of distinct polymorphs.

{\bf Relative abundance of polymorphs.} Up to this point we have characterised the emerging assemblies solely by the number of parallel and non-parallel bonds of the single particles. As this measure separates well between particles within crystalline aggregates and particles bonded in disordered domains, we use it as the basis for crystal structure detection.
To do so, we initially determine the bonding pattern characteristic of the crystal structure under consideration. Subsequently, we classify a particle as part of this particular crystalline environment if the particle itself and at least one neighbour display the specific bonding pattern associated with the crystal. As an example, we consider the P3 lattice of dmo-as1, where the local P3-bonding pattern consists of three parallel bonds. In this case, a particle is considered  P3-crystalline if the particle itself and at least one bonded neighbour have three parallel bonds. For all other crystals, the structure identification is in principle analogous, and we refer to the SI for details. 

To determine the system-wide prevalence of a particular crystal structure, we compute the ratio of the number of particles detected to be of this crystal type to the total number of bonded particles. This yields a value ranging from 0, indicating the complete absence of this specific crystal structure in the assembly, to 1, signifying that all bonded particles in the simulation box are part of this specific crystal type.
This metric allows us to compare the prevalence of different polymorphs within a single assembly and, when averaged for each state point, across all state points.
We summarize the results of the crystal structure analysis for all dmo-systems in the heat maps presented in Figure ~\ref{fig:crystals_all}e-g, along with simulation snapshots showcasing the three most prevalent crystal structures.  In the heat maps the relative prevalence of the polymorphs is represented by a single color, using a barycentric color space (depicted in the lower right corner of the heat map). As each dmo system has three dominant polymorphs, we define a barycentric triangle whose edge points are associated with the three polymorphs, where, for ease of reading, the corner colors match the colors of the respective crystals in the snapshots. 

We note that this analysis is fully automated, from the identification of the crystalline environments to the plotting of the heat map with barycentric coloring method. 
If a state point is colored in one of the corner colors of the barycentric triangle it indicates that one of the three crystals dominates over the other two. Conversely, if a state point exhibits a more mixed assembly, colors in between are used.

We start our discussion with the dominant crystals of the dmo-as1 system (see Figure~\ref{fig:crystals_all}a1-a3). The quantitative results of this analysis are presented in the heat map in Figure~\ref{fig:crystals_all}e. In this heat map only state points with an overall crystallinity above 0.05 are shown in color, while state points below this threshold are grayed out. This analysis reveals that dmo-as1 displays significant crystallinity for packing fractions above 0.1, where the extent in temperature of the crystalline region increases with higher packing, from including only $T=0.1-0.11$ for $\phi=0.1$  to including the whole upper range of temperatures from $T=0.10-0.16$ for the highest packing fraction of 0.525. 
With the barycentric  color map at hand, we clearly observe that while for lower temperatures and packing fractions, mostly Z1 and P3 crystallites are competing (as indicated by yellow/orange), the P3 lattice becomes more prevalent and eventually emerges as the dominant crystal structure at the highest temperatures and packing fractions (indicated by burgundi). 

By analyzing the colored snapshots for dmo-s1 (see the examples reported in Figure~\ref{fig:crystals_all}b1-b2) we identify a  fully bonded parallel lattice (P4) and two  partially bonded zigzag lattices (Z1 and Z2). 
It is worth stressing that, in contrast to all other three particle types, for dmo-s1 full bonding is possible without overlaps or energetic penalties. Nonetheless, bonding of all four patches results in a certain degree of strain, as the patches are almost out of each other's interaction range and barely reach to bond (see Figure~\ref{fig:strained_particle_types}h for an illustration of bond strain). The other two prevalent lattices are partially bonded: while at first sight they seem equivalent, Z1 has a 2-np-1p bonding pattern -- bonding to three neighbours --  while Z2 displays two alternating bonding patterns, where in one pattern the particles bond to two neighbours in a non-parallel fashion (2-np), while in the other pattern the particles bond to four neighbours with two non-parallel and two parallel bonds (2-np-2p). On the level of lattice geometry, these bonding pattern differences show in that, while Z1 connects the zigzagging rows with vertically alternating p-bonds, Z2 connects them with vertically aligned p-bonds. From the aspect of chirality/number of effective components, the Z1 of dmo-s1 is a racemic co-crystal with the same enantiomer arrangement as the Z1 of dmo-as1, while Z2 is an effective symmetric binary lattice of two- and four-patch particles. That said, it must be noted that Z1 and Z2 are separated only by a horizontal shift in bonding between the zigzag rows.  Additionally, Z1 and Z2 are compatible with each other and often co-occur as can be also observed in the snapshots in Figure~\ref{fig:crystals_all}b2. 

Analogously to dmo-as1, also for dmo-s1 we conduct the crystal cluster analysis resulting in the automatic plotting of the heat map that quantifies the relative abundance of the dominant crystals (see  Figure~\ref{fig:crystals_all}f). Inspecting the region of overall crystallinity in the heat map, we find the crystalline region to be at intermediate to high temperatures/packing fractions, similar to dmo-as1, with the  distinction that it extends a little further to lower densities (up to $\phi=0.05$ $versus$ $\phi=0.1$ for dmo-as1). 
When comparing the prevalence of the three dominant lattices, we find that, at low $T$ and $\phi$, P4 clearly dominates over Z1 and Z2, while, as we move to higher $T$ and/or $\phi$ there is an intermediate region where P4 competes with Z2 and -- to a lesser degree -- with Z1. Finally, for higher $T$ and $\phi$, Z2 dominates with some degree of Z1 present for some state points.  
The described heat map shows the counter intuitive observation that the close-packed and fully bonded P4 crystal is prevalent at lower packing, while the porous and partially bonded Z2 lattice prevails at high-packing. We rationalize this observation by noting that, for lower packing fractions, the crystalline region resides also at lower temperatures, where the drive to fully bond all available patches is larger (despite the unavoidable strain). For higher packing fractions the crystal region is found at higher temperatures, meaning that the drive for strained full-bonding is reduced and thus the Z2 lattice prevails. 

In dmo-s2, all emerging crystals exhibit partial bonding with three bonds per particle (see Figure~\ref{fig:crystals_all}c1). The P3 lattice (particles in burgundi) is nearly identical to the P3 lattice found in dmo-as1, with the fine distinction that the dangling patch is always a red one. As it varies which of the two red patches is dangling, the bonding pattern still yields two patch bond enantiomers, leading to an equivalent solid solution, with again two distinct pore orientations in the assembly. The PZ lattice features a bonding pattern that alternates between particles with 3-p bonds (in burgundi) and particles with 2-p and 1-np bonds (in blue), making it an effective binary mixture of two particles with different patch arrangement. In both particle bonding patterns, it is always a red patch that is left dangling. The PZ geometry can be viewed as a lattice where parallel rows of p-bonded (alternately burgundi and blue) particles are alternately connected to each other with p- and np-bonds. 
The third dominant  lattice for dmo-s2 is Z3, where all particles 2-p and 1-np bonds. Z3 lattice connects parallel rows of p-bonded particles with np-bonds only. 
Looking at the dominance across all state points, we find that, while P3 dominates most of the state points, for some state points we observe a substantial amount of PZ and Z3 environments -- highlighted by pink (more PZ) and blue (more Z3) shades.
In the dmo-s2, the crystalline region follows the same trends as the other dmo-systems, with the crystalline region extending from only including a small intermediate $T$-range at low $\phi$ to encompassing the whole upper $T$-range for the highest $\phi$. We note that, with respect to the previous dmo systems, the crystalline region is reduced in density, while its stays very similar in temperature: it starts at $\phi=0.2$ ($versus$ $\phi=0.1$ for dmo-as1 $\phi=0.05$ for dmo-s1) and at $T=0.11$ ($versus$ $T=0.1$ for both dmo-as1 and dmo-s2). As already observed with dmo-as1, for the highest packing fraction, the system clearly exhibits the P3 lattice. 
For dma-as1, two crystalline domains are observed (see the SI for details) but, as the overall crystallinity remains below $5\%$, we exclude these systems from further analysis. The visual inspection of the coloured snapshots reveals that the high np-bonding observed in Figure~\ref{fig:average_bonding} is not due to an emergent crystalline order but rather to the abundance of finite-sized, three particle clusters, embedded in a disordered network (see the SI for details). 
Finally, we evaluate the absolute amount of crystallinity, $\langle \xi \rangle$, as the ratio between the number of particles within any crystalline structure (according to the structure detection algorithm) and the number of bonded particles, averaged according to the standard statistics procedure (see Methods section). 
Interestingly, the overall crystallinity remains low over an extensive simulation time: for most systems, $\langle \xi \rangle$ is highest at $20\%$, as reported in Figure~\ref{fig:crystal_state_diagram}(a1-d1). Note that this a conservative estimate, as we do not include the domain boundaries of the crystallites. Nonetheless, the low degree of crystallinity raises the question of which mechanism hinders the crystal growth, whether it is a matter of stability or instead due to the competition between the frustrated polymorphs~\cite{lenz2017geometrical,serafin2021frustrated,tyukodi2022thermodynamic,carpenter2020heterogeneous}.  Investigations beyond the scope of this paper would be needed to gain insight into this issue. 

\section{Discussion}

\begin{figure}
\includegraphics[width=\columnwidth]{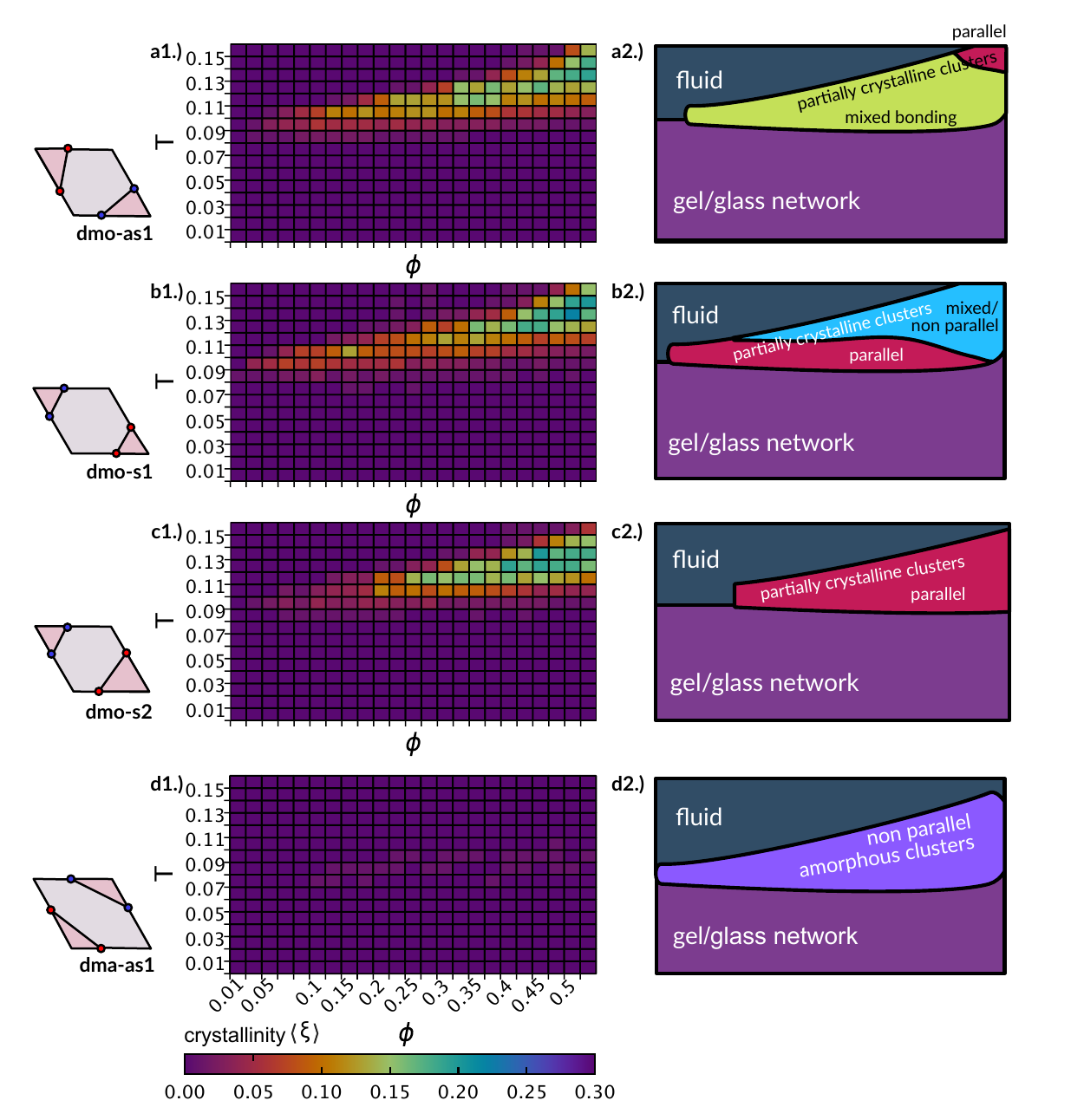}
\caption{{\bf Absolute abundance of crystal polymorphs (left) and overview of the investigated state diagrams (right).} Heat maps (a1-d1) for the overall average crystallinity, $\langle \xi \rangle$, for all studied particle types across all state points (with $\phi=0.01-0.525$ $T=0.01-0.16$), and sketches (a2-d2) for the dynamic state diagrams. Crystallinity heat map for (a1)  dmo-as1, (b1) dmo-s1, (c1) dmo-s2, (d1) dma-as1. Dynamic state diagram for (a2) dmo-as1, (b2) dmo-s1, (c2) dmo-s2, (d2) dma-as1.}
\label{fig:crystal_state_diagram}
\end{figure}

While design and inverse design strategies rely on optimizing the particles shape, functionalization and bonding geometry in order to access complex mesoscopic phases, we purposefully introduce geometric frustration in systems of building blocks designed to otherwise assemble in well-defined monolayers.  
In contrast to the guiding principle of the most recent approaches to target specific structures, which is to satisfy all possible bonds between the assembling units, we deliberately disfavour such a condition by introducing a finite set of possible full bonding failures: particle overlaps, bonding strain, bonding  incompatibility and energetic penalties. When partial bonding is favoured, disordered aggregates often emerge as a result of geometric frustration. Nonetheless, our investigation clearly shows a different route to crystallization where geometric frustration can be utilized to target complex structures.  

We summarize our results in a dynamic state diagram, reported in Figure~\ref{fig:crystal_state_diagram}(a2-d2). In the state diagram of all investigated systems, we observe a pocket of state points -- between the low-temperature disordered network region and the high-temperature/low-density fluid -- where particles have a high-bonding degree. In this region, we observe the emergence of crystallites with different, but energetically equivalent, bonding patterns for all dmo systems. Nonetheless, the assembly scenarios of those systems have some non-negligible differences with respect to each other. 

For dmo-as1, we observe both parallel and non-parallel bonding in the whole high-bonding region, emerging from the tight competition between two (out of three) porous polymorphs, where the parallel lattice eventually prevails over the zigzag one only at high temperatures and densities. In contrast, for dmo-s1, we observe that the region of high bonding has a prevalence of parallel bonding at low temperatures and densities, related to the prevalence of the only close-packed, fully bonded (but strained) parallel lattice, and a region of mixed parallel and non-parallel bonding at higher  temperatures and densities, due to the prevalence of two zigzag competing crystals. Finally, for dmo-s2, we observe a prevalence of parallel bonding, emerging from the competition between three porous polymorphs, across the whole domain of crystallinity. In contrast to those systems, for dma-as1, the amount of crystallinity in the region of high-bonding degree is very low and thus the strong non-parallel pattern features of the assemblies must be attributed to the formation of a large number of finite three-particle clusters, connected within a disordered network. 

Our results show that a different paradigm can be leveraged to build up ordered porous structures from patchy colloids, where the full bondedness of the particles is not anymore a requirement for crystallinity. The same paradigm can be used to gain insight into molecular tilings. While extended monolayers of small organic molecules have been reproduced by fully bonded, two-dimensional lattices of patchy platelets~\cite{Karner-Nanolett2019, Whitelam2012, Science_2008}, we show that the presence of dangling bonds can  produce extended crystals, suggesting that geometric frustration could be used as a guiding principle to gain a deeper understanding of organic chemistry mechanisms. Our thorough analysis of the crystal polymorphs in presence of geometric frustration highlights the emergence of chiral building units, leading to solid solutions, racemic crystals and homochiral crystals. In fact, the proposed paradigm based on polymorphism, chirality and dangling bonds play a key role in the design of bioactive organic compounds, for instance in industrial applications, where  homochiral (or enantiopure) crystals are in high demand~\cite{bernstein2020polymorphism, cruz2015facts, lee2011crystal,
calcaterra2018market, nguyen2006chiral,smith2009chiral, carpenter2020heterogeneous}.

\section{Methods}
{\bf Particle model.}
For this self-assembly study we consider regular hard rhombi with four localized patches -- one per edge -- of two distinct types, where pairs of patches are of the same type and where same-type patches attract and different-type patches repel each other. This type of building block assembles into a plethora of distinct porous and close-packed assemblies upon varying the exact placement of the four patches on the rhombus edges~\cite{Karner-Nanolett2019,karner2020patchiness,karner2024anisotropic}. In this work we focus on a subset of systems, where the bonding is geometrically frustrated: we deliberately select four patch layouts where all four patches cannot be simultaneously bonded to neighbouring particles without incurring overlaps, energetic penalties or at least a certain degree of strain. 
All  four particle types are shown in Figure~\ref{fig:strained_particle_types}(a-d), together with example motifs where the full bonding fails in Figure~\ref{fig:strained_particle_types}(g-j). The particle naming scheme leans on previous nomenclature from Ref.~\cite{Karner-Nanolett2019},  where dmo refers  to systems where the patches of the same type enclose the vertex of the small rhombus angle (Figure~\ref{fig:strained_particle_types}a-c) and dma (Figure~\ref{fig:strained_particle_types}d) refers to systems  where the patches of the same type enclose the vertex of the large  rhombus angle. 

Furthermore, the suffixes (as) and (s) denote whether same type patches are arranged asymmetrically (as) or symmetrically (s) with respect to the enclosed reference vertex, while the numbers 1 and 2 just serve as identifiers for the different arrangements.  We call the combination of these three identifiers a patch topology,  hence the four patch topologies of this study are dmo-as1, dmo-s1, dmo-s2 and dma-as1. 
The last parameter required to uniquely characterize each patch arrangement is the distance of the patches from the edge center $\Delta_c$,  which takes values between 0 (patches are in the center) and 0.5 (patches are at the vertices).  In other words, $\Delta_c$ can be viewed as a measure of patch anisotropy, that indicates how much off-center the patches are placed --  or in terms of assembly scenarios --  how far away the systems are from close-packing at $\Delta_c=0$~\cite{Karner-Nanolett2019}. While full bonding is disfavored for all off-center $\Delta_c$  with $0<\Delta_{c}<0.5$ for all four patch topologies~\cite{karner2024anisotropic}, we select $\Delta_c=0.1$ for this investigation to highlight the impact a slight patch anisotropy  on the self-assembly.

The interaction between two rhombic platelets $i$ and $j$ is characterized by a hard-core potential, 
\[ U(\vec{r}_{ij}, \Omega_{i}, \Omega_{j})  =
  \begin{cases}
    0     & \quad \text{if  $i$ and $j$ do not overlap}\\
    \infty  & \quad \text{if $i$ and $j$ do overlap}\\
  \end{cases}
\]
where $\vec{r}_{ij}$ represents the center-to-center distance vector, and $\Omega_{i}$ and $\Omega_{j}$ denote particle orientations. Overlaps between two rhombi are determined via the separating axis theorem~\cite{Golshtein_1996}. The side length of the rhombi, $l_{r}$, is set to 1.
Patch-to-patch interactions are defined by an attractive or repulsive square-well potential:
\[ W(\vec{p}_{ij})  =
  \begin{cases}
    \pm \epsilon     & \quad \text{if}\quad p_{ij}< 2r_{p}\\
    0 & \quad  \text{if}\quad p_{ij} \geq 2r_{p}, \\
  \end{cases}
\]
where $p_{ij}$ represents the patch-to-patch distance, $2r_{p}$ is the patch diameter, and $\epsilon$ denotes the patch interaction strength. We set $r_p=0.05$ to ensure that each patch can bond at maximum to another patch~\cite{karner2020matter} and $\epsilon=\pm 1$. 

{\bf Simulation details.}
We investigate the self-assembly of these rhombic building blocks using large scale real-space Monte Carlo simulations in two dimensions incorporating single particle rotation and translation moves, alongside cluster moves~\cite{Whitelam2007, Whitelam2010}. With a fixed particle count of $N=1500$, a grid of state points was constructed: $m=16$ temperatures ($T_0=0.01$ to $T_m=0.16$) and $n=24$ packing fractions ($\phi_0=0.01$ to $\phi_n=0.525$), totaling $384$ state points per system. The initial state for all systems is a square lattice of rhombi, that is melted/equilibrated for $1\times10^6$ Monte Carlo sweeps by switching off the patch interactions, which is equivalent to simulating at infinite temperature.  The self-assembly process is then started by switching on the patch interactions and setting the temperature according to the value of the respective state point. For consistency, we use the same simulation protocol for all systems across various states, performing a substantial total of $24576$ separate simulations considering all $16$ parallel runs and $384$ state points across all $4$ particle topologies.
Expecting the slow formation of disordered particle networks we conduct long simulations extending to $3-4\times 10^7$ Monte Carlo sweeps. The statistics for each state point in temperature/packing fraction is aggregated by including the last 100 checkpoints of all 16 parallel runs, where checkpoints are recorded every 10000 sweeps. Within this work, we refer to this choice of averaging as standard statistics procedure. 

\section{Acknowledgements}
The authors have used the high performance computing facilities of the Vienna Scientific Cluster(VSC) for this work.This research was funded in whole by the Austrian Science Fund (FWF) Y1163-N27. For open access purposes, the author has applied a CC BY public copyright license to any author-accepted manuscript version arising from this submission.

\section{Supporting Information}
\subsection{Crystal structure identification} 
\begin{figure*}
\includegraphics[width=0.6\textwidth]{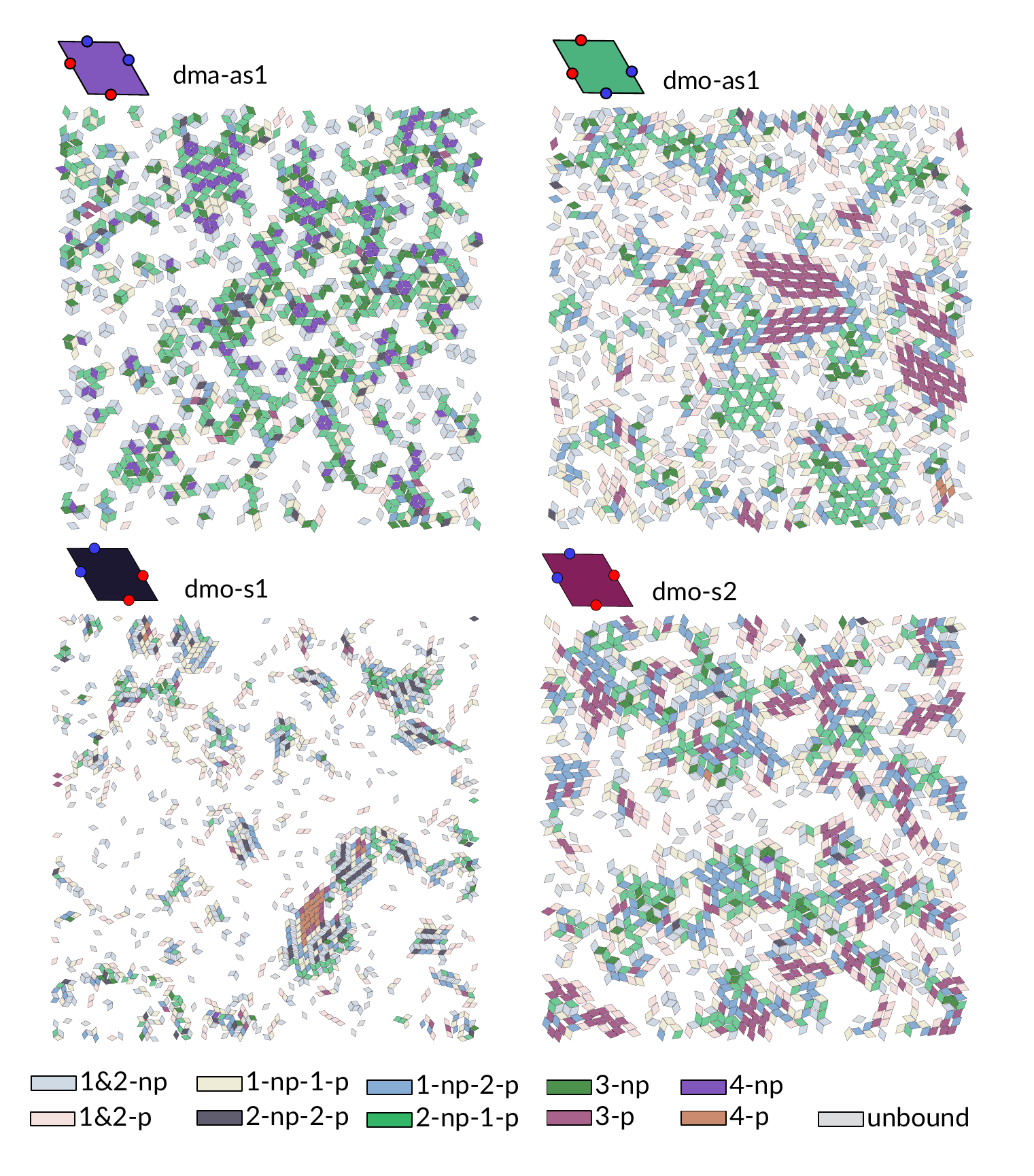}
\caption{\small{{\bf Bonded neighbourhoods in sample snapshots of the four investigated systems.} As labelled: dma-as1 ($\phi=0.45$, $T=0.13$), dmo-as1 ($\phi=0.5$, $T=0.14$), dmo-s1 ($\phi=0.225$, $T=0.12$) and dmo-s2 ($\phi=0.5$, $T=0.13$). Generally label abbreviations follow the scheme a-np-b-p, where a is the number of non-parallel bonds and b the number of parallel bonds. For example,  2-np-2-p denotes a particle with two non-parallel and two parallel bonds. Exceptions to this scheme are $1$\&2-np and $1$\&2-p, that indicate particles with either one or two non-parallel (np) or one or two parallel (p) bonds.}}
\label{fig:neighbourhood}
\end{figure*}

The four patch topologies featured in our work -- dma-as1, dmo-as1, dmo-s1 and dmo-s2 -- all give rise to a set of crystal polymorphs with distinct symmetries. 
\\
{\bf Bonding pattern identification.} We find that these polymorphs can be most effectively distinguished from each other via their bonding patterns. In this case we define a bonding pattern to be the number of parallel (p) and non-parallel(np) bonded neighbours of the particle.
After the visual inspection of all available snapshots, we find in total ten bonding patterns that contribute to the formation of crystalline units.
In Figure~\ref{fig:neighbourhood} we show sample snapshots for all four systems, where the particles are algorithmically colored according to these neighbourhoods, where we list all considered bonding patterns in the legend. It is important to note that, for up to three bonded neighbours, the listed bonding patterns encompass all bonding possibilities. In contrast, for four bonds only four parallel bonds (4-p), four non-parallel bonds (4-np) and two parallel plus two non-parallel bonds (2-np-2p) are observed, while bonding patterns with three parallel (non-parallel) bonds and one non-parallel (parallel) bond are never observed as part of a crystallite.  
Due to the coloring we can already observe, in all systems, some of the crystallites with their typical bonding patterns and symmetries visually distinguished from the liquid around, as the number of bonds 
is on average three while the average bonding for the liquid is between one and three. 
By inspecting the snapshots, however, we can also see that not all bonded neighbours have a crystal symmetry. Therefore the next step must be a crystal structure identification that robustly identifies whether a particle is bonded within a crystal neighbourhood and not bonded by chance -- at the single particle level -- according to a specific bonding pattern. 
\\
{\bf From bonding patterns to crystal structures.} In general, for a particle to be seen as part of a crystalline neighbourhood, the particles neighbours have to be arranged in a certain symmetry around the particle, and, additionally the neighbours themselves have to be subject to a neighbourhood of the same crystalline symmetry. 
Ordinarily, one would therefore rely on symmetry detecting positional order parameters such as the Steinhardt bond order parameters, that take into account distance vectors of the nearest neighbours for crystal structure detection  \cite{steinhardt1983bond}. 
In this work, however, we refrain from using the Steinhardt order parameters, because they rely purely on center-to-center positions, while our crystals exhibit a complex orientational ordering.
Global structure identification methods, such as radial distribution function or structure factor, would prove insufficient because of the presence of competing polymorphs. This means that we would not be able to assign emerging crystal peaks to a specific crystal structure. 
As we realized that in all but one system -- the dmo-as1 system we will discuss later --- the set of bonding patterns are unique to each polymorph, we built our crystal detection on top of the highlighted bonding patterns. Specifically, from visual inspection we identified by hand 11 different sets of bonding patterns (in total) as candidates for the crystal structure identification. Here it must be added that, while in principle, it is possible to overlook crystal structures via visual inspection, the highlighting due to the coloring of the bonding patterns and the fact that we thoroughly inspected the outcomes of all simulation runs, re-assures us that we did not overlook pre-dominantly occurring crystal structures. 
\begin{figure*}
\includegraphics[width=0.6\textwidth]{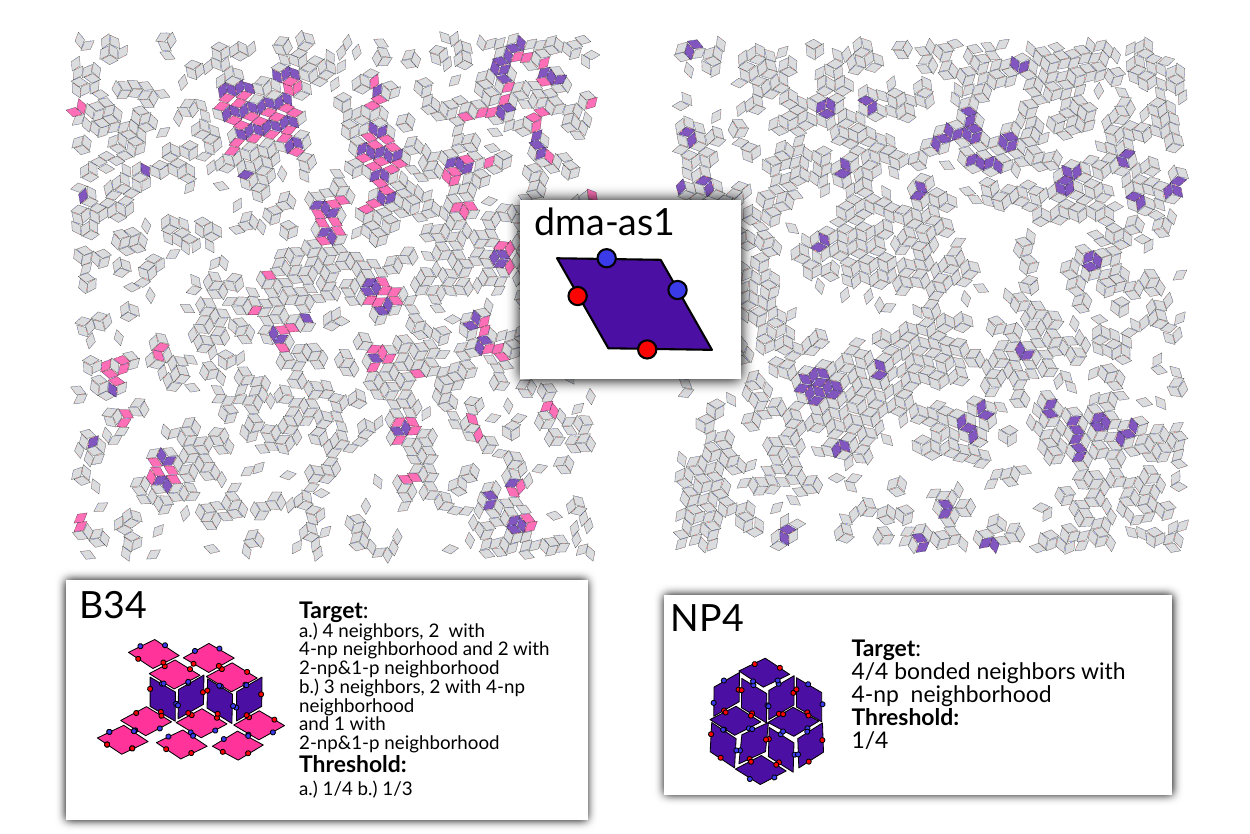}
\caption{\small{ {\bf Crystal structure detection for dma-as1.} On the left: simulation snapshots at $\phi=0.45$, $T=0.13$ with particles detected as B34 particles colored: particles with 4-np bonding pattern in lilac, particles with 2-np-1p in pink. Inset: Target and threshold for B34 structure detection. On the right: simulation snapshots at $\phi=0.5$, $T=0.13$} with only particles detected as NP4 crystal colord in lilac. Inset: Target and threshold for NP4 structure detection.}
\label{fig:dmaas1}
\end{figure*}
To illustrate how the structure identification process is set up we go through the crystal candidate algorithm for dma-as1. The insets of Figure~\ref{fig:dmaas1} show sketches of the two identified crystal candidates. The first is the B34 crystal, where zigzagging rows of 4-np particles (in lilac) are connected via particles with 2-np bonds and 1-p bond (in pink), and the second is the NP4 crystal, where the particles are fully bonded with np-bonds (4-np), creating a close-packed, non-parallel monolayer. 
To classify a particle as crystalline, we essentially follow a pattern matching algorithm, where we define the bonding pattern of the perfect target crystal as well as a threshold to account for crystal defects. 
Target pattern and threshold are given for B34 and NP4 in the inset of Figure~\ref{fig:dmaas1}, next to the crystal sketches. In the perfect target B34, crystal particles can be in one of two environments. In the first one (case a. in the inset of Figure~\ref{fig:dmaas1}), the particle itself is 4-np bonded, and two neighbours must be 4-np bonded themselves and two neighbours must be 2-np-1p bonded in order for the particle to be classified as B34.  In the other case (case b. in the inset of Figure ~\ref{fig:dmaas1}), the particle is itself 2-np-1-p bonded, then two neighbours must 4-np bonded and one neighbour must be 2-np-1-p bonded. 
To allow for a degree of imperfections we set threshold values. To be classified as B34-crystalline a particle must be either (a) 4-np bonded or (b) 2-np-1p bonded and in both cases at least one of the bonded neighbours must be either 2-np-1p bonded or 4-np bonded. In the case (a) that is one out of four bonded neighbours -- denoted as 1/4 in the legend of Figure~\ref{fig:dmaas1} --- or one out of three bonded neighbours (1/3) in case (b). The target and threshold instructions for the NP4 crystal can be read equivalently to the B34 structure. For NP4, there is only one target pattern: a particle is defined as NP4-crystalline if the particle itself and all four bonded neighbours are bonded 4-np.  Analogously, the threshold requirement is that at least the particle itself and one neighbour is 4-np bonded (1/4).

To show how well this crystal detection performs, we added two simulation snapshots in Figure~\ref{fig:dmaas1}, where we use the reported target and threshold to automatically identify B34 environments (on the left) and NP4 environments (on the right). We find that, while the larger crystallites are identified correctly, for smaller clusters we observe some mismatches. We solve this problem (for dma-as1 and in general for all other systems) by labeling a particle as crystalline only if it is within a crystal environment larger than 5 crystalline particles.
Note that a particle can be a member of two crystal structures, as illustrated in Figure~\ref{fig:dmaas1}, where with the chosen threshold, the 4-np row (in lilac) within the B34 which is both part of B34 and also detected as NP4. We regard this as a feature because it enables us to identify the crystal structure of single defect lines, but it can be in principle alleviated with a stricter threshold. 
Note, however, that to estimate the total crystallinity $\xi$ -- i.e. the total number of all crystalline particles over all bonded particles -- we only count whether a particle is crystalline or not and do not double count in cases where a particle is classified as part of two crystal structures. 

Overall, the snapshots in Figure~\ref{fig:dmaas1} illustrate what the automatic crystal structure detection across all simulations runs found: despite the presence of some small B34 and NP4 crystallites larger than 5 particles, the overall crystallinity is low.
In fact our calculation over all simulation runs show that the average crystallinity $\langle \xi\rangle$ of dma-as1 is below the threshold of 5\% for all temperature and density state points (see Figure~5 in the main text) and we conclude that the self assembly products of dma-as1 are not crystalline in character. 

\begin{figure*}
\includegraphics[width=0.6\textwidth]{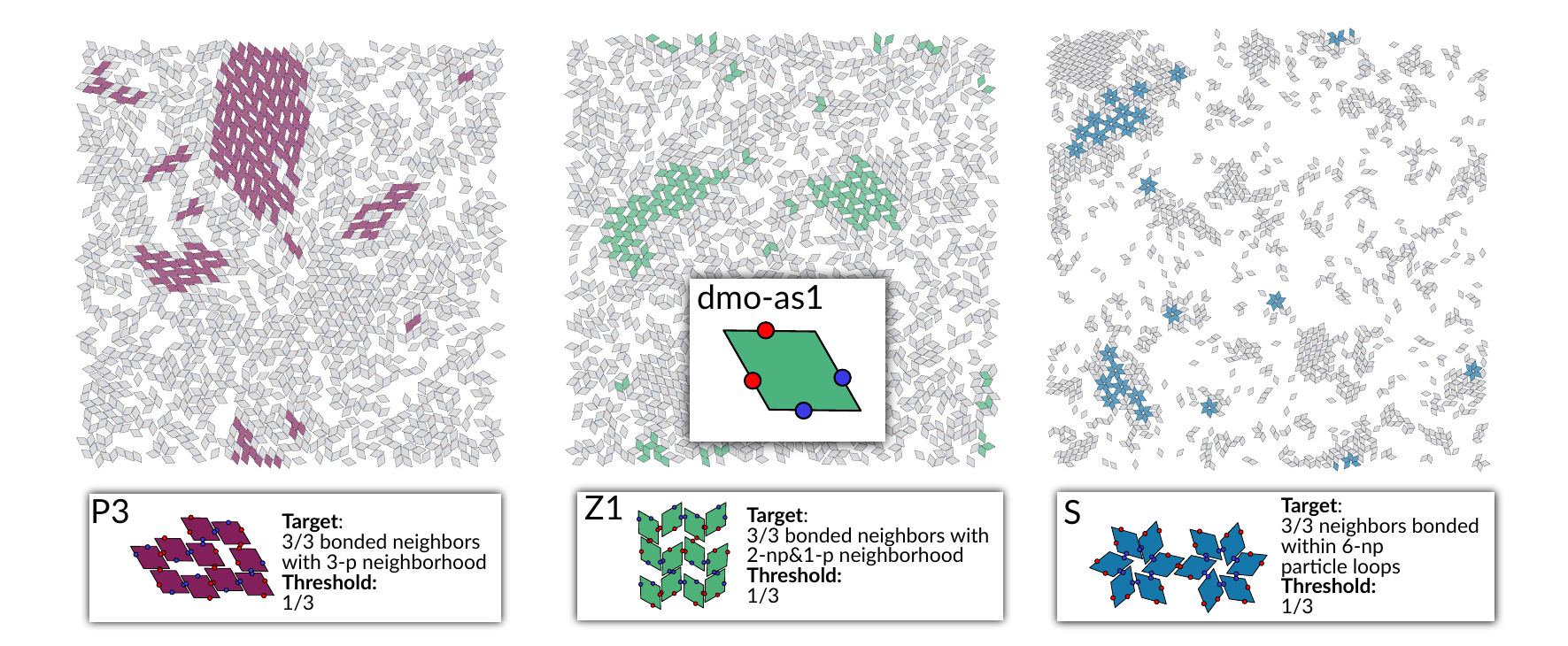}
\caption{\small{{\bf Crystal structure detection for dmo-as1.} On the left: simulation snapshots at $\phi=0.525$, $T=0.15$ with particles detected as P3 particles colored in red. Inset: target and threshold for P3 structure detection. In the middle: simulation snapshots at $\phi=0.475$, $T=0.15$ with particles detected as Z1 colored in green. Inset: target and threshold for Z1 structure detection. On the right: simulation snapshots at $\phi=0.225$, $T=0.12$ with particles detected as S colored in blue. Inset: target and threshold for S structure detection.}}
\label{fig:dmoas1}
\end{figure*}

\begin{figure*}
\includegraphics[width=0.6\textwidth]{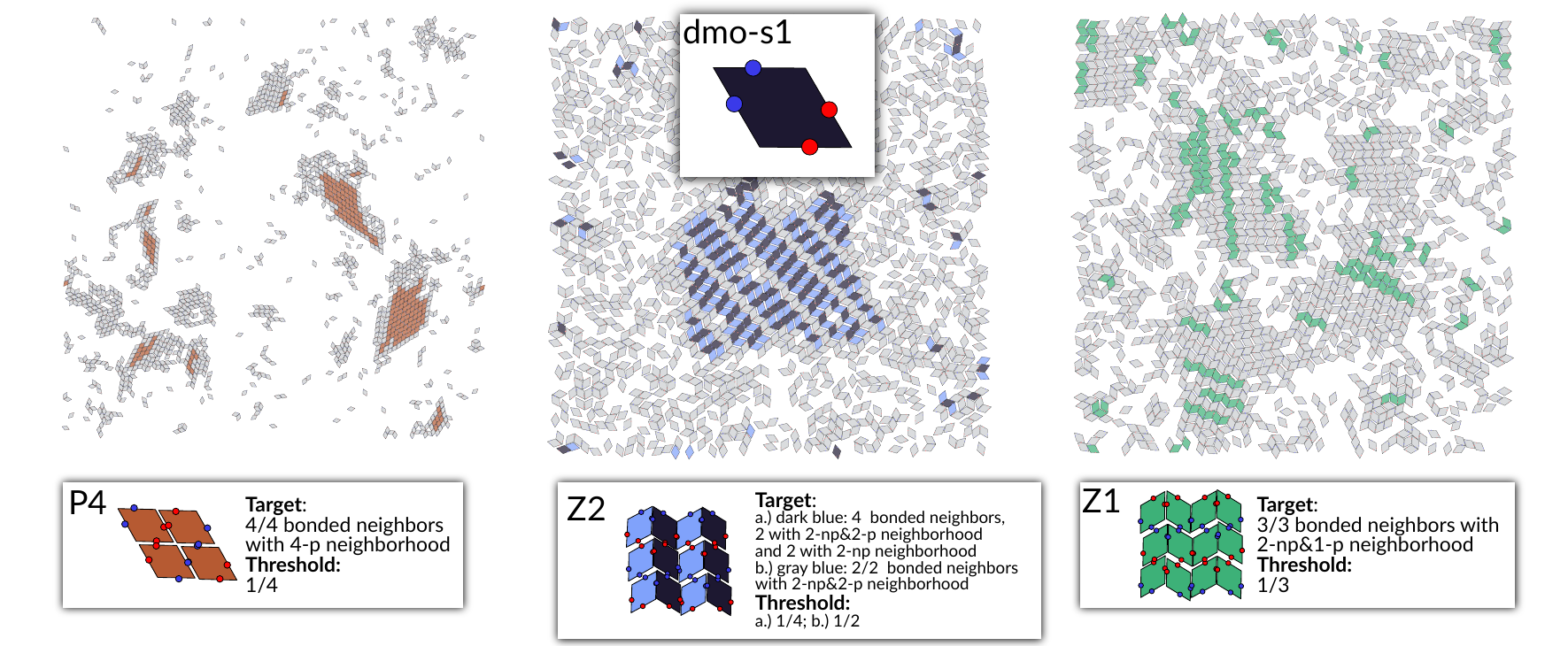}
\caption{\small{{\bf Crystal structure detection for dmo-s1.} On the left: simulation snapshots at $\phi=0.15$, $T=0.11$ with particles detected as P4 particles colored in brown. Inset: target and threshold for P4 structure detection. In the middle: simulation snapshots at $\phi=0.5$, $T=0.16$ with particles detected as Z2 colored in black (2-np-2p bonding pattern) and light blue (2-np bonding pattern). Inset: target and threshold for Z2 structure detection. On the right: simulation snapshots at $\phi=0.5$, $T=0.14$ with particles detected as Z1 colored in green. Inset: target and threshold for Z1 structure detection.}}
\label{fig:dmos1}
\end{figure*}

\begin{figure*}
\includegraphics[width=0.6\textwidth]{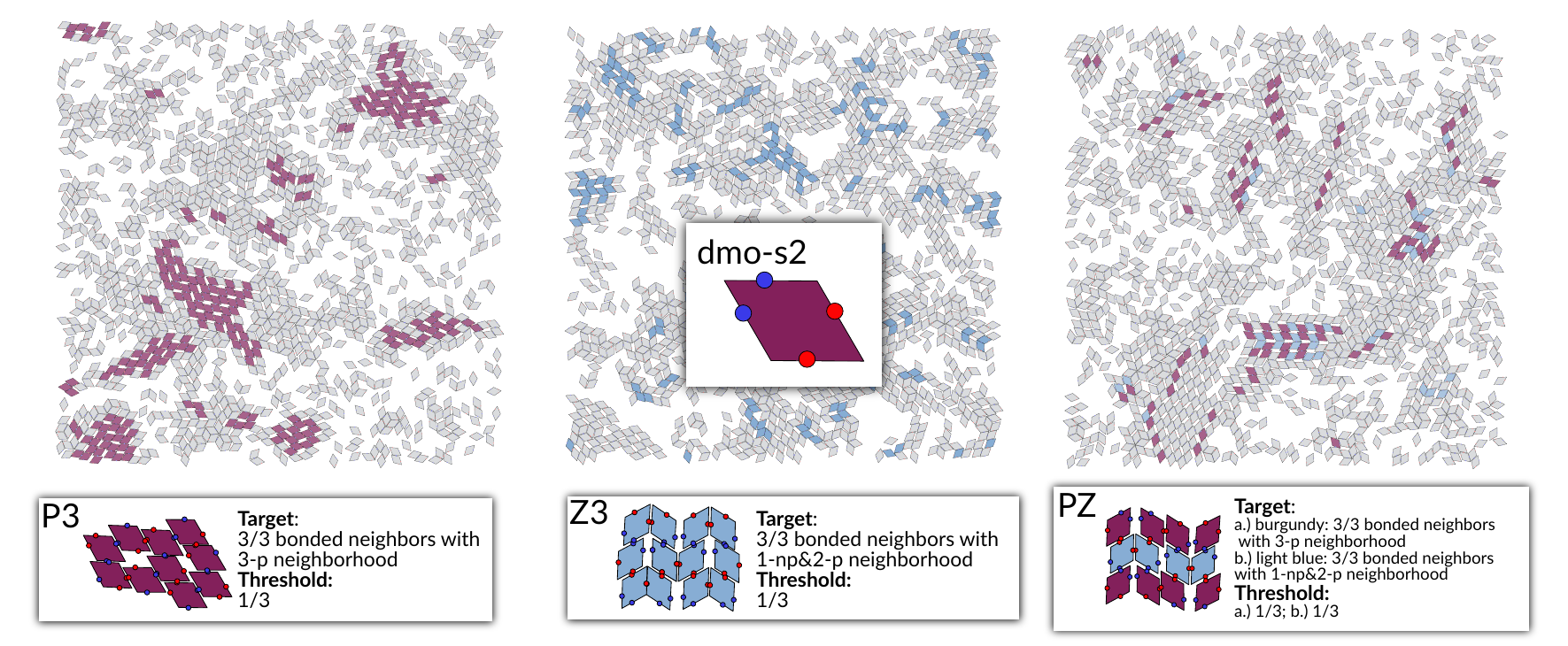}
\caption{\small{{\bf Crystal structure detection for dmo-s1.} On the left: simulation snapshots at $\phi=0.475$, $T=0.14$ with particles detected as P3 particles colored in red. Inset: target and threshold for P3 structure detection. In the middle: simulation snapshots at $\phi=0.5$, $T=0.13$ with particles detected as Z3 colored in blue. Inset: target and threshold for Z3 structure detection. On the right: simulation snapshots at $\phi=0.5$, $T=0.14$ with particles detected as PZ colored in red (3-p bonding pattern) and blue (1-np-2p bonding pattern). Inset: target and threshold for PZ structure detection.}}
\label{fig:dmos2}
\end{figure*}

For the other three systems - dmo-as1, dmo-s1 and dmo-s2 -- the target and threshold instructions can be read in the same way as for B34 and NP4. 
For the dmo-as1 systems, shown in Figure~\ref{fig:dmoas1}, where we find larger crystallites present, particularly for the P3 and the Z1 crystals, the performance of the crystal detection algorithm can be judged more thoroughly. At first, the threshold requirement, that only the particle itself and one neighbour has to adhere to the target crystal structure, may seem weak. However, by visual inspection of the P3 and Z1 systems, we find that the structure detection is in fact rather accurate, an in contrast, turns out to be even quite strict, as boundary particles are not classified as crystalline in most cases, as the particles themselves do not fulfill the bonding pattern - either because they miss bonds or because one or more bonds is mismatched with respect to the required bonding pattern.  
For the crystal structure detection of the S crystal in the dmo-as1 (in Figure~\ref{fig:dmoas1} on the right) a caveat must be added: the Z1 lattice and the S lattice are the only identified lattices with the same bonding pattern -- two non-parallel bonds and one parallel bond. In this exceptional case, we distinguish the S-particles from the Z1-particles by additionally checking if the particle is bonded within a non-parallel loop of size six (6-np loop). 
Once the bonding patterns have been determined, the crystal structure detection works analogously, where the target structure is that all three (3/3) bonded neighbours are bonded within a 6-np loop, while the threshold is that at least one of the neighbours is bonded in a 6-np loop (1/3). 
The detected crystal structures for the dmo-s1 systems -- P4, Z2 and Z1 -- in Figure ~\ref{fig:dmos1} further confirm that the crystal identification via bonding patterns is very accurate and rather strict, as can be seen especially well in the fully bonded P4 clusters, where all particles at the boundary of the cluster are excluded due to the fact that they are not fully bonded. 
For the Z2 system (Figure~\ref{fig:dmos1} in the center) we can observe that the defect rows are also excluded form the Z2 crystallite. We note that the defect lines themselves adhere to an Z1-bonding pattern and are detected as such by the algorithm, as illustrated -- albeit by a different snapshot -- in Figure~\ref{fig:dmos1} on the right hand side. 
In dmo-s1, Z1-crystalline rows are in fact most often seen as defect lines within Z2, and it might be interesting to investigate in the future if the Z1-crystal for dmo-s1 only grows as a defect of Z2.  
At this point, we note and reiterate, that it is due to the chosen low crystallinity threshold, that we can distinguish the crystal symmetry of a single defect row and thus reason about it. 
Finally, we show the detection of the crystal polymorph of dmo-s2 in Figure~\ref{fig:dmos2}: P3, Z3 and PZ.  While P3 in Figure~\ref{fig:dmos2} on the left , and to a lesser extent PZ in Figure~\ref{fig:dmos2} on the right,  exhibit clearly detectable crystallites, there are many small scattered Z3 environments without a clear crystal symmetry visible, highlighting the importance of maintaining minimum threshold of -- in this case five -- bonded crystalline particles. 

%


\clearpage
\begin{thebibliography}{62}%
\makeatletter
\providecommand \@ifxundefined [1]{%
 \@ifx{#1\undefined}
}%
\providecommand \@ifnum [1]{%
 \ifnum #1\expandafter \@firstoftwo
 \else \expandafter \@secondoftwo
 \fi
}%
\providecommand \@ifx [1]{%
 \ifx #1\expandafter \@firstoftwo
 \else \expandafter \@secondoftwo
 \fi
}%
\providecommand \natexlab [1]{#1}%
\providecommand \enquote  [1]{``#1''}%
\providecommand \bibnamefont  [1]{#1}%
\providecommand \bibfnamefont [1]{#1}%
\providecommand \citenamefont [1]{#1}%
\providecommand \href@noop [0]{\@secondoftwo}%
\providecommand \href [0]{\begingroup \@sanitize@url \@href}%
\providecommand \@href[1]{\@@startlink{#1}\@@href}%
\providecommand \@@href[1]{\endgroup#1\@@endlink}%
\providecommand \@sanitize@url [0]{\catcode `\\12\catcode `\$12\catcode
  `\&12\catcode `\#12\catcode `\^12\catcode `\_12\catcode `\%12\relax}%
\providecommand \@@startlink[1]{}%
\providecommand \@@endlink[0]{}%
\providecommand \url  [0]{\begingroup\@sanitize@url \@url }%
\providecommand \@url [1]{\endgroup\@href {#1}{\urlprefix }}%
\providecommand \urlprefix  [0]{URL }%
\providecommand \Eprint [0]{\href }%
\providecommand \doibase [0]{https://doi.org/}%
\providecommand \selectlanguage [0]{\@gobble}%
\providecommand \bibinfo  [0]{\@secondoftwo}%
\providecommand \bibfield  [0]{\@secondoftwo}%
\providecommand \translation [1]{[#1]}%
\providecommand \BibitemOpen [0]{}%
\providecommand \bibitemStop [0]{}%
\providecommand \bibitemNoStop [0]{.\EOS\space}%
\providecommand \EOS [0]{\spacefactor3000\relax}%
\providecommand \BibitemShut  [1]{\csname bibitem#1\endcsname}%
\let\auto@bib@innerbib\@empty
\bibitem [{\citenamefont {Pawar}\ and\ \citenamefont
  {Kretzschmar}(2010)}]{pawar2010fabrication}%
  \BibitemOpen
  \bibfield  {author} {\bibinfo {author} {\bibfnamefont {A.~B.}\ \bibnamefont
  {Pawar}}\ and\ \bibinfo {author} {\bibfnamefont {I.}~\bibnamefont
  {Kretzschmar}},\ }\bibfield  {title} {\bibinfo {title} {Fabrication,
  assembly, and application of patchy particles},\ }\href@noop {} {\bibfield
  {journal} {\bibinfo  {journal} {Macromolecular rapid communications}\
  }\textbf {\bibinfo {volume} {31}},\ \bibinfo {pages} {150} (\bibinfo {year}
  {2010})}\BibitemShut {NoStop}%
\bibitem [{\citenamefont {Bianchi}\ \emph {et~al.}(2011)\citenamefont
  {Bianchi}, \citenamefont {Blaak},\ and\ \citenamefont
  {Likos}}]{bianchi2011patchy}%
  \BibitemOpen
  \bibfield  {author} {\bibinfo {author} {\bibfnamefont {E.}~\bibnamefont
  {Bianchi}}, \bibinfo {author} {\bibfnamefont {R.}~\bibnamefont {Blaak}},\
  and\ \bibinfo {author} {\bibfnamefont {C.~N.}\ \bibnamefont {Likos}},\
  }\bibfield  {title} {\bibinfo {title} {Patchy colloids: state of the art and
  perspectives},\ }\href@noop {} {\bibfield  {journal} {\bibinfo  {journal}
  {Physical Chemistry Chemical Physics}\ }\textbf {\bibinfo {volume} {13}},\
  \bibinfo {pages} {6397} (\bibinfo {year} {2011})}\BibitemShut {NoStop}%
\bibitem [{\citenamefont {Sacanna}\ and\ \citenamefont
  {Pine}(2011)}]{Sacanna2011}%
  \BibitemOpen
  \bibfield  {author} {\bibinfo {author} {\bibfnamefont {S.}~\bibnamefont
  {Sacanna}}\ and\ \bibinfo {author} {\bibfnamefont {D.~J.}\ \bibnamefont
  {Pine}},\ }\bibfield  {title} {\bibinfo {title} {{Shape-anisotropic colloids:
  Building blocks for complex assemblies}},\ }\href@noop {} {\bibfield
  {journal} {\bibinfo  {journal} {Curr. Opin. Colloid Interface Sci.}\ }\textbf
  {\bibinfo {volume} {16}},\ \bibinfo {pages} {96} (\bibinfo {year}
  {2011})}\BibitemShut {NoStop}%
\bibitem [{\citenamefont {Walther}\ and\ \citenamefont
  {Muller}(2013)}]{walther2013janus}%
  \BibitemOpen
  \bibfield  {author} {\bibinfo {author} {\bibfnamefont {A.}~\bibnamefont
  {Walther}}\ and\ \bibinfo {author} {\bibfnamefont {A.~H.}\ \bibnamefont
  {Muller}},\ }\bibfield  {title} {\bibinfo {title} {Janus particles:
  synthesis, self-assembly, physical properties, and applications},\
  }\href@noop {} {\bibfield  {journal} {\bibinfo  {journal} {Chemical reviews}\
  }\textbf {\bibinfo {volume} {113}},\ \bibinfo {pages} {5194} (\bibinfo {year}
  {2013})}\BibitemShut {NoStop}%
\bibitem [{\citenamefont {Boles}\ \emph {et~al.}(2016)\citenamefont {Boles},
  \citenamefont {Engel},\ and\ \citenamefont {Talapin}}]{Boles2016}%
  \BibitemOpen
  \bibfield  {author} {\bibinfo {author} {\bibfnamefont {M.~A.}\ \bibnamefont
  {Boles}}, \bibinfo {author} {\bibfnamefont {M.}~\bibnamefont {Engel}},\ and\
  \bibinfo {author} {\bibfnamefont {D.~V.}\ \bibnamefont {Talapin}},\
  }\bibfield  {title} {\bibinfo {title} {Self-assembly of colloidal
  nanocrystals: From intricate structures to functional materials},\
  }\href@noop {} {\bibfield  {journal} {\bibinfo  {journal} {Chemical Reviews}\
  }\textbf {\bibinfo {volume} {116}},\ \bibinfo {pages} {11220} (\bibinfo
  {year} {2016})}\BibitemShut {NoStop}%
\bibitem [{\citenamefont {Xu}\ \emph {et~al.}(2018)\citenamefont {Xu},
  \citenamefont {Li},\ and\ \citenamefont {Yin}}]{Small_2018}%
  \BibitemOpen
  \bibfield  {author} {\bibinfo {author} {\bibfnamefont {W.}~\bibnamefont
  {Xu}}, \bibinfo {author} {\bibfnamefont {Z.}~\bibnamefont {Li}},\ and\
  \bibinfo {author} {\bibfnamefont {Y.}~\bibnamefont {Yin}},\ }\bibfield
  {title} {\bibinfo {title} {{Colloidal Assembly Approaches to
  Micro/Nanostructures of Complex Morphologies}},\ }\href@noop {} {\bibfield
  {journal} {\bibinfo  {journal} {Small}\ }\textbf {\bibinfo {volume} {14}},\
  \bibinfo {pages} {1801083} (\bibinfo {year} {2018})}\BibitemShut {NoStop}%
\bibitem [{\citenamefont {D{\'e}sert}\ \emph {et~al.}(2013)\citenamefont
  {D{\'e}sert}, \citenamefont {Hubert}, \citenamefont {Fu}, \citenamefont
  {Moulet}, \citenamefont {Majimel}, \citenamefont {Barboteau}, \citenamefont
  {Thill}, \citenamefont {Lansalot}, \citenamefont {Bourgeat-Lami},
  \citenamefont {Duguet},\ and\ \citenamefont {Ravaine}}]{Ravaine2013}%
  \BibitemOpen
  \bibfield  {author} {\bibinfo {author} {\bibfnamefont {A.}~\bibnamefont
  {D{\'e}sert}}, \bibinfo {author} {\bibfnamefont {C.}~\bibnamefont {Hubert}},
  \bibinfo {author} {\bibfnamefont {Z.}~\bibnamefont {Fu}}, \bibinfo {author}
  {\bibfnamefont {L.}~\bibnamefont {Moulet}}, \bibinfo {author} {\bibfnamefont
  {J.}~\bibnamefont {Majimel}}, \bibinfo {author} {\bibfnamefont
  {P.}~\bibnamefont {Barboteau}}, \bibinfo {author} {\bibfnamefont
  {A.}~\bibnamefont {Thill}}, \bibinfo {author} {\bibfnamefont
  {M.}~\bibnamefont {Lansalot}}, \bibinfo {author} {\bibfnamefont
  {E.}~\bibnamefont {Bourgeat-Lami}}, \bibinfo {author} {\bibfnamefont
  {E.}~\bibnamefont {Duguet}},\ and\ \bibinfo {author} {\bibfnamefont
  {S.}~\bibnamefont {Ravaine}},\ }\bibfield  {title} {\bibinfo {title}
  {{Synthesis and Site-Specific Functionalization of Tetravalent, Hexavalent,
  and Dodecavalent Silica Particles}},\ }\href@noop {} {\bibfield  {journal}
  {\bibinfo  {journal} {Angew. Chem., Int. Ed.}\ }\textbf {\bibinfo {volume}
  {52}},\ \bibinfo {pages} {11068} (\bibinfo {year} {2013})}\BibitemShut
  {NoStop}%
\bibitem [{\citenamefont {Sacanna}\ \emph {et~al.}(2013)\citenamefont
  {Sacanna}, \citenamefont {Korpics}, \citenamefont {Rodriguez}, \citenamefont
  {Col{\'o}n-Mel{\'e}ndez}, \citenamefont {Kim}, \citenamefont {Pine},\ and\
  \citenamefont {Yi}}]{sacanna2013shaping}%
  \BibitemOpen
  \bibfield  {author} {\bibinfo {author} {\bibfnamefont {S.}~\bibnamefont
  {Sacanna}}, \bibinfo {author} {\bibfnamefont {M.}~\bibnamefont {Korpics}},
  \bibinfo {author} {\bibfnamefont {K.}~\bibnamefont {Rodriguez}}, \bibinfo
  {author} {\bibfnamefont {L.}~\bibnamefont {Col{\'o}n-Mel{\'e}ndez}}, \bibinfo
  {author} {\bibfnamefont {S.-H.}\ \bibnamefont {Kim}}, \bibinfo {author}
  {\bibfnamefont {D.~J.}\ \bibnamefont {Pine}},\ and\ \bibinfo {author}
  {\bibfnamefont {G.-R.}\ \bibnamefont {Yi}},\ }\bibfield  {title} {\bibinfo
  {title} {Shaping colloids for self-assembly},\ }\href@noop {} {\bibfield
  {journal} {\bibinfo  {journal} {Nature communications}\ }\textbf {\bibinfo
  {volume} {4}},\ \bibinfo {pages} {1688} (\bibinfo {year} {2013})}\BibitemShut
  {NoStop}%
\bibitem [{\citenamefont {Ravaine}\ and\ \citenamefont
  {Duguet}(2017)}]{ravaine2017synthesis}%
  \BibitemOpen
  \bibfield  {author} {\bibinfo {author} {\bibfnamefont {S.}~\bibnamefont
  {Ravaine}}\ and\ \bibinfo {author} {\bibfnamefont {E.}~\bibnamefont
  {Duguet}},\ }\bibfield  {title} {\bibinfo {title} {Synthesis and assembly of
  patchy particles: Recent progress and future prospects},\ }\href@noop {}
  {\bibfield  {journal} {\bibinfo  {journal} {Current opinion in colloid \&
  interface science}\ }\textbf {\bibinfo {volume} {30}},\ \bibinfo {pages} {45}
  (\bibinfo {year} {2017})}\BibitemShut {NoStop}%
\bibitem [{\citenamefont {Wang}\ \emph {et~al.}(2012)\citenamefont {Wang},
  \citenamefont {Wang}, \citenamefont {Breed}, \citenamefont {Manoharan},
  \citenamefont {Feng}, \citenamefont {Hollingsworth}, \citenamefont {Weck},\
  and\ \citenamefont {Pine}}]{wang2012colloids}%
  \BibitemOpen
  \bibfield  {author} {\bibinfo {author} {\bibfnamefont {Y.}~\bibnamefont
  {Wang}}, \bibinfo {author} {\bibfnamefont {Y.}~\bibnamefont {Wang}}, \bibinfo
  {author} {\bibfnamefont {D.~R.}\ \bibnamefont {Breed}}, \bibinfo {author}
  {\bibfnamefont {V.~N.}\ \bibnamefont {Manoharan}}, \bibinfo {author}
  {\bibfnamefont {L.}~\bibnamefont {Feng}}, \bibinfo {author} {\bibfnamefont
  {A.~D.}\ \bibnamefont {Hollingsworth}}, \bibinfo {author} {\bibfnamefont
  {M.}~\bibnamefont {Weck}},\ and\ \bibinfo {author} {\bibfnamefont {D.~J.}\
  \bibnamefont {Pine}},\ }\bibfield  {title} {\bibinfo {title} {Colloids with
  valence and specific directional bonding},\ }\href@noop {} {\bibfield
  {journal} {\bibinfo  {journal} {Nature}\ }\textbf {\bibinfo {volume} {491}},\
  \bibinfo {pages} {51} (\bibinfo {year} {2012})}\BibitemShut {NoStop}%
\bibitem [{\citenamefont {He}\ \emph {et~al.}(2020)\citenamefont {He},
  \citenamefont {Gales}, \citenamefont {Ducrot}, \citenamefont {Gong},
  \citenamefont {Yi}, \citenamefont {Sacanna},\ and\ \citenamefont
  {Pine}}]{he2020colloidal}%
  \BibitemOpen
  \bibfield  {author} {\bibinfo {author} {\bibfnamefont {M.}~\bibnamefont
  {He}}, \bibinfo {author} {\bibfnamefont {J.~P.}\ \bibnamefont {Gales}},
  \bibinfo {author} {\bibfnamefont {{\'E}.}~\bibnamefont {Ducrot}}, \bibinfo
  {author} {\bibfnamefont {Z.}~\bibnamefont {Gong}}, \bibinfo {author}
  {\bibfnamefont {G.-R.}\ \bibnamefont {Yi}}, \bibinfo {author} {\bibfnamefont
  {S.}~\bibnamefont {Sacanna}},\ and\ \bibinfo {author} {\bibfnamefont {D.~J.}\
  \bibnamefont {Pine}},\ }\bibfield  {title} {\bibinfo {title} {Colloidal
  diamond},\ }\href@noop {} {\bibfield  {journal} {\bibinfo  {journal}
  {Nature}\ }\textbf {\bibinfo {volume} {585}},\ \bibinfo {pages} {524}
  (\bibinfo {year} {2020})}\BibitemShut {NoStop}%
\bibitem [{\citenamefont {Chen}\ \emph {et~al.}(2011)\citenamefont {Chen},
  \citenamefont {Bae},\ and\ \citenamefont {Granick}}]{chen2011directed}%
  \BibitemOpen
  \bibfield  {author} {\bibinfo {author} {\bibfnamefont {Q.}~\bibnamefont
  {Chen}}, \bibinfo {author} {\bibfnamefont {S.~C.}\ \bibnamefont {Bae}},\ and\
  \bibinfo {author} {\bibfnamefont {S.}~\bibnamefont {Granick}},\ }\bibfield
  {title} {\bibinfo {title} {Directed self-assembly of a colloidal kagome
  lattice},\ }\href@noop {} {\bibfield  {journal} {\bibinfo  {journal}
  {Nature}\ }\textbf {\bibinfo {volume} {469}},\ \bibinfo {pages} {381}
  (\bibinfo {year} {2011})}\BibitemShut {NoStop}%
\bibitem [{\citenamefont {Diaz~A}\ \emph {et~al.}(2020)\citenamefont {Diaz~A},
  \citenamefont {Oh}, \citenamefont {Yi},\ and\ \citenamefont
  {Pine}}]{diaz2020photo}%
  \BibitemOpen
  \bibfield  {author} {\bibinfo {author} {\bibfnamefont {J.~A.}\ \bibnamefont
  {Diaz~A}}, \bibinfo {author} {\bibfnamefont {J.~S.}\ \bibnamefont {Oh}},
  \bibinfo {author} {\bibfnamefont {G.-R.}\ \bibnamefont {Yi}},\ and\ \bibinfo
  {author} {\bibfnamefont {D.~J.}\ \bibnamefont {Pine}},\ }\bibfield  {title}
  {\bibinfo {title} {Photo-printing of faceted {DNA} patchy particles},\
  }\href@noop {} {\bibfield  {journal} {\bibinfo  {journal} {Proceedings of the
  National Academy of Sciences}\ }\textbf {\bibinfo {volume} {117}},\ \bibinfo
  {pages} {10645} (\bibinfo {year} {2020})}\BibitemShut {NoStop}%
\bibitem [{\citenamefont {Oh}\ \emph {et~al.}(2019)\citenamefont {Oh},
  \citenamefont {Lee}, \citenamefont {Glotzer}, \citenamefont {Yi},\ and\
  \citenamefont {Pine}}]{oh2019colloidal}%
  \BibitemOpen
  \bibfield  {author} {\bibinfo {author} {\bibfnamefont {J.~S.}\ \bibnamefont
  {Oh}}, \bibinfo {author} {\bibfnamefont {S.}~\bibnamefont {Lee}}, \bibinfo
  {author} {\bibfnamefont {S.~C.}\ \bibnamefont {Glotzer}}, \bibinfo {author}
  {\bibfnamefont {G.-R.}\ \bibnamefont {Yi}},\ and\ \bibinfo {author}
  {\bibfnamefont {D.~J.}\ \bibnamefont {Pine}},\ }\bibfield  {title} {\bibinfo
  {title} {Colloidal fibers and rings by cooperative assembly},\ }\href@noop {}
  {\bibfield  {journal} {\bibinfo  {journal} {Nature communications}\ }\textbf
  {\bibinfo {volume} {10}},\ \bibinfo {pages} {3936} (\bibinfo {year}
  {2019})}\BibitemShut {NoStop}%
\bibitem [{\citenamefont {Zhang}\ \emph {et~al.}(2017)\citenamefont {Zhang},
  \citenamefont {Grzybowski},\ and\ \citenamefont {Granick}}]{zhang2017janus}%
  \BibitemOpen
  \bibfield  {author} {\bibinfo {author} {\bibfnamefont {J.}~\bibnamefont
  {Zhang}}, \bibinfo {author} {\bibfnamefont {B.~A.}\ \bibnamefont
  {Grzybowski}},\ and\ \bibinfo {author} {\bibfnamefont {S.}~\bibnamefont
  {Granick}},\ }\bibfield  {title} {\bibinfo {title} {Janus particle synthesis,
  assembly, and application},\ }\href@noop {} {\bibfield  {journal} {\bibinfo
  {journal} {Langmuir}\ }\textbf {\bibinfo {volume} {33}},\ \bibinfo {pages}
  {6964} (\bibinfo {year} {2017})}\BibitemShut {NoStop}%
\bibitem [{\citenamefont {Liu}\ \emph {et~al.}(2016)\citenamefont {Liu},
  \citenamefont {Halverson}, \citenamefont {Tian}, \citenamefont {Tkachenko},\
  and\ \citenamefont {Gang}}]{liu2016self}%
  \BibitemOpen
  \bibfield  {author} {\bibinfo {author} {\bibfnamefont {W.}~\bibnamefont
  {Liu}}, \bibinfo {author} {\bibfnamefont {J.}~\bibnamefont {Halverson}},
  \bibinfo {author} {\bibfnamefont {Y.}~\bibnamefont {Tian}}, \bibinfo {author}
  {\bibfnamefont {A.~V.}\ \bibnamefont {Tkachenko}},\ and\ \bibinfo {author}
  {\bibfnamefont {O.}~\bibnamefont {Gang}},\ }\bibfield  {title} {\bibinfo
  {title} {Self-organized architectures from assorted {DNA}-framed
  nanoparticles},\ }\href@noop {} {\bibfield  {journal} {\bibinfo  {journal}
  {Nature chemistry}\ }\textbf {\bibinfo {volume} {8}},\ \bibinfo {pages} {867}
  (\bibinfo {year} {2016})}\BibitemShut {NoStop}%
\bibitem [{\citenamefont {Klinkova}\ \emph {et~al.}(2013)\citenamefont
  {Klinkova}, \citenamefont {Th{\'e}rien-Aubin}, \citenamefont {Choueiri},
  \citenamefont {Rubinstein},\ and\ \citenamefont
  {Kumacheva}}]{klinkova2013colloidal}%
  \BibitemOpen
  \bibfield  {author} {\bibinfo {author} {\bibfnamefont {A.}~\bibnamefont
  {Klinkova}}, \bibinfo {author} {\bibfnamefont {H.}~\bibnamefont
  {Th{\'e}rien-Aubin}}, \bibinfo {author} {\bibfnamefont {R.~M.}\ \bibnamefont
  {Choueiri}}, \bibinfo {author} {\bibfnamefont {M.}~\bibnamefont
  {Rubinstein}},\ and\ \bibinfo {author} {\bibfnamefont {E.}~\bibnamefont
  {Kumacheva}},\ }\bibfield  {title} {\bibinfo {title} {Colloidal analogs of
  molecular chain stoppers},\ }\href@noop {} {\bibfield  {journal} {\bibinfo
  {journal} {Proceedings of the National Academy of Sciences}\ }\textbf
  {\bibinfo {volume} {110}},\ \bibinfo {pages} {18775} (\bibinfo {year}
  {2013})}\BibitemShut {NoStop}%
\bibitem [{\citenamefont {Iubini}\ \emph {et~al.}(2020)\citenamefont {Iubini},
  \citenamefont {Baiesi},\ and\ \citenamefont {Orlandini}}]{iubini2020aging}%
  \BibitemOpen
  \bibfield  {author} {\bibinfo {author} {\bibfnamefont {S.}~\bibnamefont
  {Iubini}}, \bibinfo {author} {\bibfnamefont {M.}~\bibnamefont {Baiesi}},\
  and\ \bibinfo {author} {\bibfnamefont {E.}~\bibnamefont {Orlandini}},\
  }\bibfield  {title} {\bibinfo {title} {Aging of living polymer networks: A
  model with patchy particles},\ }\href@noop {} {\bibfield  {journal} {\bibinfo
   {journal} {Soft Matter}\ }\textbf {\bibinfo {volume} {16}},\ \bibinfo
  {pages} {9543} (\bibinfo {year} {2020})}\BibitemShut {NoStop}%
\bibitem [{\citenamefont {Sciortino}\ \emph {et~al.}(2007)\citenamefont
  {Sciortino}, \citenamefont {Bianchi}, \citenamefont {Douglas},\ and\
  \citenamefont {Tartaglia}}]{sciortino2007self}%
  \BibitemOpen
  \bibfield  {author} {\bibinfo {author} {\bibfnamefont {F.}~\bibnamefont
  {Sciortino}}, \bibinfo {author} {\bibfnamefont {E.}~\bibnamefont {Bianchi}},
  \bibinfo {author} {\bibfnamefont {J.~F.}\ \bibnamefont {Douglas}},\ and\
  \bibinfo {author} {\bibfnamefont {P.}~\bibnamefont {Tartaglia}},\ }\bibfield
  {title} {\bibinfo {title} {Self-assembly of patchy particles into polymer
  chains: A parameter-free comparison between wertheim theory and {Monte Carlo}
  simulation},\ }\href@noop {} {\bibfield  {journal} {\bibinfo  {journal} {The
  Journal of chemical physics}\ }\textbf {\bibinfo {volume} {126}} (\bibinfo
  {year} {2007})}\BibitemShut {NoStop}%
\bibitem [{\citenamefont {Doppelbauer}\ \emph {et~al.}(2010)\citenamefont
  {Doppelbauer}, \citenamefont {Bianchi},\ and\ \citenamefont
  {Kahl}}]{doppelbauer2010self}%
  \BibitemOpen
  \bibfield  {author} {\bibinfo {author} {\bibfnamefont {G.}~\bibnamefont
  {Doppelbauer}}, \bibinfo {author} {\bibfnamefont {E.}~\bibnamefont
  {Bianchi}},\ and\ \bibinfo {author} {\bibfnamefont {G.}~\bibnamefont
  {Kahl}},\ }\bibfield  {title} {\bibinfo {title} {Self-assembly scenarios of
  patchy colloidal particles in two dimensions},\ }\href@noop {} {\bibfield
  {journal} {\bibinfo  {journal} {Journal of Physics: Condensed Matter}\
  }\textbf {\bibinfo {volume} {22}},\ \bibinfo {pages} {104105} (\bibinfo
  {year} {2010})}\BibitemShut {NoStop}%
\bibitem [{\citenamefont {Doye}\ \emph {et~al.}(2007)\citenamefont {Doye},
  \citenamefont {Louis}, \citenamefont {Lin}, \citenamefont {Allen},
  \citenamefont {Noya}, \citenamefont {Wilber}, \citenamefont {Kok},\ and\
  \citenamefont {Lyus}}]{doye2007controlling}%
  \BibitemOpen
  \bibfield  {author} {\bibinfo {author} {\bibfnamefont {J.~P.}\ \bibnamefont
  {Doye}}, \bibinfo {author} {\bibfnamefont {A.~A.}\ \bibnamefont {Louis}},
  \bibinfo {author} {\bibfnamefont {I.-C.}\ \bibnamefont {Lin}}, \bibinfo
  {author} {\bibfnamefont {L.~R.}\ \bibnamefont {Allen}}, \bibinfo {author}
  {\bibfnamefont {E.~G.}\ \bibnamefont {Noya}}, \bibinfo {author}
  {\bibfnamefont {A.~W.}\ \bibnamefont {Wilber}}, \bibinfo {author}
  {\bibfnamefont {H.~C.}\ \bibnamefont {Kok}},\ and\ \bibinfo {author}
  {\bibfnamefont {R.}~\bibnamefont {Lyus}},\ }\bibfield  {title} {\bibinfo
  {title} {Controlling crystallization and its absence: proteins, colloids and
  patchy models},\ }\href@noop {} {\bibfield  {journal} {\bibinfo  {journal}
  {Physical Chemistry Chemical Physics}\ }\textbf {\bibinfo {volume} {9}},\
  \bibinfo {pages} {2197} (\bibinfo {year} {2007})}\BibitemShut {NoStop}%
\bibitem [{\citenamefont {Reinhardt}\ \emph {et~al.}(2013)\citenamefont
  {Reinhardt}, \citenamefont {Romano},\ and\ \citenamefont
  {Doye}}]{reinhardt2013computing}%
  \BibitemOpen
  \bibfield  {author} {\bibinfo {author} {\bibfnamefont {A.}~\bibnamefont
  {Reinhardt}}, \bibinfo {author} {\bibfnamefont {F.}~\bibnamefont {Romano}},\
  and\ \bibinfo {author} {\bibfnamefont {J.~P.}\ \bibnamefont {Doye}},\
  }\bibfield  {title} {\bibinfo {title} {Computing phase diagrams for a
  quasicrystal-forming patchy-particle system},\ }\href@noop {} {\bibfield
  {journal} {\bibinfo  {journal} {Physical review letters}\ }\textbf {\bibinfo
  {volume} {110}},\ \bibinfo {pages} {255503} (\bibinfo {year}
  {2013})}\BibitemShut {NoStop}%
\bibitem [{\citenamefont {Liu}\ \emph {et~al.}(2019)\citenamefont {Liu},
  \citenamefont {Li}, \citenamefont {Li},\ and\ \citenamefont
  {Mao}}]{liu2019rational}%
  \BibitemOpen
  \bibfield  {author} {\bibinfo {author} {\bibfnamefont {L.}~\bibnamefont
  {Liu}}, \bibinfo {author} {\bibfnamefont {Z.}~\bibnamefont {Li}}, \bibinfo
  {author} {\bibfnamefont {Y.}~\bibnamefont {Li}},\ and\ \bibinfo {author}
  {\bibfnamefont {C.}~\bibnamefont {Mao}},\ }\bibfield  {title} {\bibinfo
  {title} {Rational design and self-assembly of two-dimensional, dodecagonal
  {DNA} quasicrystals},\ }\href@noop {} {\bibfield  {journal} {\bibinfo
  {journal} {Journal of the American Chemical Society}\ }\textbf {\bibinfo
  {volume} {141}},\ \bibinfo {pages} {4248} (\bibinfo {year}
  {2019})}\BibitemShut {NoStop}%
\bibitem [{\citenamefont {Morphew}\ \emph {et~al.}(2018)\citenamefont
  {Morphew}, \citenamefont {Shaw}, \citenamefont {Avins},\ and\ \citenamefont
  {Chakrabarti}}]{morphew2018programming}%
  \BibitemOpen
  \bibfield  {author} {\bibinfo {author} {\bibfnamefont {D.}~\bibnamefont
  {Morphew}}, \bibinfo {author} {\bibfnamefont {J.}~\bibnamefont {Shaw}},
  \bibinfo {author} {\bibfnamefont {C.}~\bibnamefont {Avins}},\ and\ \bibinfo
  {author} {\bibfnamefont {D.}~\bibnamefont {Chakrabarti}},\ }\bibfield
  {title} {\bibinfo {title} {Programming hierarchical self-assembly of patchy
  particles into colloidal crystals via colloidal molecules},\ }\href@noop {}
  {\bibfield  {journal} {\bibinfo  {journal} {ACS nano}\ }\textbf {\bibinfo
  {volume} {12}},\ \bibinfo {pages} {2355} (\bibinfo {year}
  {2018})}\BibitemShut {NoStop}%
\bibitem [{\citenamefont {Ye}\ \emph {et~al.}(2013)\citenamefont {Ye},
  \citenamefont {Chen}, \citenamefont {Engel}, \citenamefont {Millan},
  \citenamefont {Li}, \citenamefont {Qi}, \citenamefont {Xing}, \citenamefont
  {Collins}, \citenamefont {Kagan}, \citenamefont {Li}, \citenamefont
  {Glotzer},\ and\ \citenamefont {Murray}}]{Ye2013}%
  \BibitemOpen
  \bibfield  {author} {\bibinfo {author} {\bibfnamefont {X.}~\bibnamefont
  {Ye}}, \bibinfo {author} {\bibfnamefont {J.}~\bibnamefont {Chen}}, \bibinfo
  {author} {\bibfnamefont {M.}~\bibnamefont {Engel}}, \bibinfo {author}
  {\bibfnamefont {J.~a.}\ \bibnamefont {Millan}}, \bibinfo {author}
  {\bibfnamefont {W.}~\bibnamefont {Li}}, \bibinfo {author} {\bibfnamefont
  {L.}~\bibnamefont {Qi}}, \bibinfo {author} {\bibfnamefont {G.}~\bibnamefont
  {Xing}}, \bibinfo {author} {\bibfnamefont {J.~E.}\ \bibnamefont {Collins}},
  \bibinfo {author} {\bibfnamefont {C.~R.}\ \bibnamefont {Kagan}}, \bibinfo
  {author} {\bibfnamefont {J.}~\bibnamefont {Li}}, \bibinfo {author}
  {\bibfnamefont {S.~C.}\ \bibnamefont {Glotzer}},\ and\ \bibinfo {author}
  {\bibfnamefont {C.~B.}\ \bibnamefont {Murray}},\ }\bibfield  {title}
  {\bibinfo {title} {{Competition of shape and interaction patchiness for
  self-assembling nanoplates}},\ }\href@noop {} {\bibfield  {journal} {\bibinfo
   {journal} {Nat. Chem.}\ }\textbf {\bibinfo {volume} {5}},\ \bibinfo {pages}
  {466} (\bibinfo {year} {2013})}\BibitemShut {NoStop}%
\bibitem [{\citenamefont {Millan}\ \emph {et~al.}(2014)\citenamefont {Millan},
  \citenamefont {Ortiz}, \citenamefont {{Van Anders}},\ and\ \citenamefont
  {Glotzer}}]{Millan2014}%
  \BibitemOpen
  \bibfield  {author} {\bibinfo {author} {\bibfnamefont {J.~A.}\ \bibnamefont
  {Millan}}, \bibinfo {author} {\bibfnamefont {D.}~\bibnamefont {Ortiz}},
  \bibinfo {author} {\bibfnamefont {G.}~\bibnamefont {{Van Anders}}},\ and\
  \bibinfo {author} {\bibfnamefont {S.~C.}\ \bibnamefont {Glotzer}},\
  }\bibfield  {title} {\bibinfo {title} {{Self-assembly of archimedean tilings
  with enthalpically and entropically patchy polygons}},\ }\href@noop {}
  {\bibfield  {journal} {\bibinfo  {journal} {ACS Nano}\ }\textbf {\bibinfo
  {volume} {8}},\ \bibinfo {pages} {2918} (\bibinfo {year} {2014})}\BibitemShut
  {NoStop}%
\bibitem [{\citenamefont {Kang}\ \emph {et~al.}(2016)\citenamefont {Kang},
  \citenamefont {Choi}, \citenamefont {Kim}, \citenamefont {Yeom},
  \citenamefont {Lee}, \citenamefont {Park},\ and\ \citenamefont
  {Lee}}]{Lee_2016}%
  \BibitemOpen
  \bibfield  {author} {\bibinfo {author} {\bibfnamefont {S.-M.}\ \bibnamefont
  {Kang}}, \bibinfo {author} {\bibfnamefont {C.-H.}\ \bibnamefont {Choi}},
  \bibinfo {author} {\bibfnamefont {J.}~\bibnamefont {Kim}}, \bibinfo {author}
  {\bibfnamefont {S.-J.}\ \bibnamefont {Yeom}}, \bibinfo {author}
  {\bibfnamefont {D.}~\bibnamefont {Lee}}, \bibinfo {author} {\bibfnamefont
  {B.~J.}\ \bibnamefont {Park}},\ and\ \bibinfo {author} {\bibfnamefont
  {C.-S.}\ \bibnamefont {Lee}},\ }\bibfield  {title} {\bibinfo {title}
  {{Capillarity-induced directed self-assembly of patchy hexagram particles at
  the air-water interface}},\ }\href@noop {} {\bibfield  {journal} {\bibinfo
  {journal} {Soft Matter}\ }\textbf {\bibinfo {volume} {12}},\ \bibinfo {pages}
  {5847} (\bibinfo {year} {2016})}\BibitemShut {NoStop}%
\bibitem [{\citenamefont {Whitelam}\ \emph {et~al.}(2012)\citenamefont
  {Whitelam}, \citenamefont {Tamblyn}, \citenamefont {Beton},\ and\
  \citenamefont {Garrahan}}]{Whitelam2012}%
  \BibitemOpen
  \bibfield  {author} {\bibinfo {author} {\bibfnamefont {S.}~\bibnamefont
  {Whitelam}}, \bibinfo {author} {\bibfnamefont {I.}~\bibnamefont {Tamblyn}},
  \bibinfo {author} {\bibfnamefont {P.~H.}\ \bibnamefont {Beton}},\ and\
  \bibinfo {author} {\bibfnamefont {J.~P.}\ \bibnamefont {Garrahan}},\
  }\bibfield  {title} {\bibinfo {title} {Random and ordered phases of
  off-lattice rhombus tiles},\ }\href@noop {} {\bibfield  {journal} {\bibinfo
  {journal} {Phys. Rev. Lett.}\ }\textbf {\bibinfo {volume} {108}},\ \bibinfo
  {pages} {035702} (\bibinfo {year} {2012})}\BibitemShut {NoStop}%
\bibitem [{\citenamefont {Karner}\ \emph {et~al.}(2019)\citenamefont {Karner},
  \citenamefont {Dellago},\ and\ \citenamefont
  {Bianchi}}]{Karner-Nanolett2019}%
  \BibitemOpen
  \bibfield  {author} {\bibinfo {author} {\bibfnamefont {C.}~\bibnamefont
  {Karner}}, \bibinfo {author} {\bibfnamefont {C.}~\bibnamefont {Dellago}},\
  and\ \bibinfo {author} {\bibfnamefont {E.}~\bibnamefont {Bianchi}},\
  }\bibfield  {title} {\bibinfo {title} {Design of patchy rhombi: from
  close-packed tilings to open lattices},\ }\href@noop {} {\bibfield  {journal}
  {\bibinfo  {journal} {Nano Letters}\ }\textbf {\bibinfo {volume} {19}},\
  \bibinfo {pages} {7806} (\bibinfo {year} {2019})}\BibitemShut {NoStop}%
\bibitem [{\citenamefont {Pakalidou}\ \emph {et~al.}(2020)\citenamefont
  {Pakalidou}, \citenamefont {Mu}, \citenamefont {Masters},\ and\ \citenamefont
  {Avenda{\~n}o}}]{pakalidou2020engineering}%
  \BibitemOpen
  \bibfield  {author} {\bibinfo {author} {\bibfnamefont {N.}~\bibnamefont
  {Pakalidou}}, \bibinfo {author} {\bibfnamefont {J.}~\bibnamefont {Mu}},
  \bibinfo {author} {\bibfnamefont {A.~J.}\ \bibnamefont {Masters}},\ and\
  \bibinfo {author} {\bibfnamefont {C.}~\bibnamefont {Avenda{\~n}o}},\
  }\bibfield  {title} {\bibinfo {title} {Engineering porous two-dimensional
  lattices via self-assembly of non-convex hexagonal platelets},\ }\href@noop
  {} {\bibfield  {journal} {\bibinfo  {journal} {Molecular Systems Design \&
  Engineering}\ }\textbf {\bibinfo {volume} {5}},\ \bibinfo {pages} {376}
  (\bibinfo {year} {2020})}\BibitemShut {NoStop}%
\bibitem [{\citenamefont {Cai}\ \emph {et~al.}(2018)\citenamefont {Cai},
  \citenamefont {Mineart}, \citenamefont {Li}, \citenamefont {Spontak},
  \citenamefont {Manners},\ and\ \citenamefont {Qiu}}]{Cai2018hierarchical}%
  \BibitemOpen
  \bibfield  {author} {\bibinfo {author} {\bibfnamefont {J.}~\bibnamefont
  {Cai}}, \bibinfo {author} {\bibfnamefont {K.~P.}\ \bibnamefont {Mineart}},
  \bibinfo {author} {\bibfnamefont {X.}~\bibnamefont {Li}}, \bibinfo {author}
  {\bibfnamefont {R.~J.}\ \bibnamefont {Spontak}}, \bibinfo {author}
  {\bibfnamefont {I.}~\bibnamefont {Manners}},\ and\ \bibinfo {author}
  {\bibfnamefont {H.}~\bibnamefont {Qiu}},\ }\bibfield  {title} {\bibinfo
  {title} {Hierarchical self-assembly of toroidal micelles into
  multidimensional nanoporous superstructures},\ }\href@noop {} {\bibfield
  {journal} {\bibinfo  {journal} {ACS Macro Letters}\ }\textbf {\bibinfo
  {volume} {7}},\ \bibinfo {pages} {1040} (\bibinfo {year} {2018})}\BibitemShut
  {NoStop}%
\bibitem [{\citenamefont {Kocabey}\ \emph {et~al.}(2015)\citenamefont
  {Kocabey}, \citenamefont {Kempter}, \citenamefont {List}, \citenamefont
  {Xing}, \citenamefont {Bae}, \citenamefont {Schiffels}, \citenamefont {Shih},
  \citenamefont {Simmel},\ and\ \citenamefont {Liedl}}]{kocabey2015membrane}%
  \BibitemOpen
  \bibfield  {author} {\bibinfo {author} {\bibfnamefont {S.}~\bibnamefont
  {Kocabey}}, \bibinfo {author} {\bibfnamefont {S.}~\bibnamefont {Kempter}},
  \bibinfo {author} {\bibfnamefont {J.}~\bibnamefont {List}}, \bibinfo {author}
  {\bibfnamefont {Y.}~\bibnamefont {Xing}}, \bibinfo {author} {\bibfnamefont
  {W.}~\bibnamefont {Bae}}, \bibinfo {author} {\bibfnamefont {D.}~\bibnamefont
  {Schiffels}}, \bibinfo {author} {\bibfnamefont {W.~M.}\ \bibnamefont {Shih}},
  \bibinfo {author} {\bibfnamefont {F.~C.}\ \bibnamefont {Simmel}},\ and\
  \bibinfo {author} {\bibfnamefont {T.}~\bibnamefont {Liedl}},\ }\bibfield
  {title} {\bibinfo {title} {Membrane-assisted growth of {DNA} origami
  nanostructure arrays},\ }\href@noop {} {\bibfield  {journal} {\bibinfo
  {journal} {ACS nano}\ }\textbf {\bibinfo {volume} {9}},\ \bibinfo {pages}
  {3530} (\bibinfo {year} {2015})}\BibitemShut {NoStop}%
\bibitem [{\citenamefont {Journot}\ \emph {et~al.}(2019)\citenamefont
  {Journot}, \citenamefont {Ramakrishna}, \citenamefont {Wallace},\ and\
  \citenamefont {Turberfield}}]{journot2019modifying}%
  \BibitemOpen
  \bibfield  {author} {\bibinfo {author} {\bibfnamefont {C.~M.}\ \bibnamefont
  {Journot}}, \bibinfo {author} {\bibfnamefont {V.}~\bibnamefont
  {Ramakrishna}}, \bibinfo {author} {\bibfnamefont {M.~I.}\ \bibnamefont
  {Wallace}},\ and\ \bibinfo {author} {\bibfnamefont {A.~J.}\ \bibnamefont
  {Turberfield}},\ }\bibfield  {title} {\bibinfo {title} {Modifying membrane
  morphology and interactions with {DNA} origami clathrin-mimic networks},\
  }\href@noop {} {\bibfield  {journal} {\bibinfo  {journal} {ACS nano}\
  }\textbf {\bibinfo {volume} {13}},\ \bibinfo {pages} {9973} (\bibinfo {year}
  {2019})}\BibitemShut {NoStop}%
\bibitem [{\citenamefont {Tikhomirov}\ \emph {et~al.}(2018)\citenamefont
  {Tikhomirov}, \citenamefont {Petersen},\ and\ \citenamefont
  {Qian}}]{Qian_2018}%
  \BibitemOpen
  \bibfield  {author} {\bibinfo {author} {\bibfnamefont {G.}~\bibnamefont
  {Tikhomirov}}, \bibinfo {author} {\bibfnamefont {P.}~\bibnamefont
  {Petersen}},\ and\ \bibinfo {author} {\bibfnamefont {L.}~\bibnamefont
  {Qian}},\ }\bibfield  {title} {\bibinfo {title} {{Triangular {DNA} Origami
  Tilings}},\ }\href@noop {} {\bibfield  {journal} {\bibinfo  {journal} {J. Am.
  Chem. Soc.}\ }\textbf {\bibinfo {volume} {140}},\ \bibinfo {pages} {17361}
  (\bibinfo {year} {2018})}\BibitemShut {NoStop}%
\bibitem [{\citenamefont {Ma}\ and\ \citenamefont
  {Ferguson}(2019)}]{ma2019inverse}%
  \BibitemOpen
  \bibfield  {author} {\bibinfo {author} {\bibfnamefont {Y.}~\bibnamefont
  {Ma}}\ and\ \bibinfo {author} {\bibfnamefont {A.~L.}\ \bibnamefont
  {Ferguson}},\ }\bibfield  {title} {\bibinfo {title} {Inverse design of
  self-assembling colloidal crystals with omnidirectional photonic bandgaps},\
  }\href@noop {} {\bibfield  {journal} {\bibinfo  {journal} {Soft matter}\
  }\textbf {\bibinfo {volume} {15}},\ \bibinfo {pages} {8808} (\bibinfo {year}
  {2019})}\BibitemShut {NoStop}%
\bibitem [{\citenamefont {Long}\ and\ \citenamefont
  {Ferguson}(2018)}]{long2018rational}%
  \BibitemOpen
  \bibfield  {author} {\bibinfo {author} {\bibfnamefont {A.~W.}\ \bibnamefont
  {Long}}\ and\ \bibinfo {author} {\bibfnamefont {A.~L.}\ \bibnamefont
  {Ferguson}},\ }\bibfield  {title} {\bibinfo {title} {Rational design of
  patchy colloids via landscape engineering},\ }\href@noop {} {\bibfield
  {journal} {\bibinfo  {journal} {Molecular Systems Design \& Engineering}\
  }\textbf {\bibinfo {volume} {3}},\ \bibinfo {pages} {49} (\bibinfo {year}
  {2018})}\BibitemShut {NoStop}%
\bibitem [{\citenamefont {Chen}\ \emph {et~al.}(2018)\citenamefont {Chen},
  \citenamefont {Zhang},\ and\ \citenamefont {Torquato}}]{chen2018inverse}%
  \BibitemOpen
  \bibfield  {author} {\bibinfo {author} {\bibfnamefont {D.}~\bibnamefont
  {Chen}}, \bibinfo {author} {\bibfnamefont {G.}~\bibnamefont {Zhang}},\ and\
  \bibinfo {author} {\bibfnamefont {S.}~\bibnamefont {Torquato}},\ }\bibfield
  {title} {\bibinfo {title} {Inverse design of colloidal crystals via optimized
  patchy interactions},\ }\href@noop {} {\bibfield  {journal} {\bibinfo
  {journal} {The Journal of Physical Chemistry B}\ }\textbf {\bibinfo {volume}
  {122}},\ \bibinfo {pages} {8462} (\bibinfo {year} {2018})}\BibitemShut
  {NoStop}%
\bibitem [{\citenamefont {Lieu}\ and\ \citenamefont
  {Yoshinaga}(2022)}]{lieu2022inverse}%
  \BibitemOpen
  \bibfield  {author} {\bibinfo {author} {\bibfnamefont {U.~T.}\ \bibnamefont
  {Lieu}}\ and\ \bibinfo {author} {\bibfnamefont {N.}~\bibnamefont
  {Yoshinaga}},\ }\bibfield  {title} {\bibinfo {title} {Inverse design of
  two-dimensional structure by self-assembly of patchy particles},\ }\href@noop
  {} {\bibfield  {journal} {\bibinfo  {journal} {The Journal of Chemical
  Physics}\ }\textbf {\bibinfo {volume} {156}} (\bibinfo {year}
  {2022})}\BibitemShut {NoStop}%
\bibitem [{\citenamefont {Rivera-Rivera}\ \emph {et~al.}(2023)\citenamefont
  {Rivera-Rivera}, \citenamefont {Moore},\ and\ \citenamefont
  {Glotzer}}]{rivera2023inverse}%
  \BibitemOpen
  \bibfield  {author} {\bibinfo {author} {\bibfnamefont {L.~Y.}\ \bibnamefont
  {Rivera-Rivera}}, \bibinfo {author} {\bibfnamefont {T.~C.}\ \bibnamefont
  {Moore}},\ and\ \bibinfo {author} {\bibfnamefont {S.~C.}\ \bibnamefont
  {Glotzer}},\ }\bibfield  {title} {\bibinfo {title} {Inverse design of
  triblock janus spheres for self-assembly of complex structures in the
  crystallization slot via digital alchemy},\ }\href@noop {} {\bibfield
  {journal} {\bibinfo  {journal} {Soft Matter}\ }\textbf {\bibinfo {volume}
  {19}},\ \bibinfo {pages} {2726} (\bibinfo {year} {2023})}\BibitemShut
  {NoStop}%
\bibitem [{\citenamefont {Truong-Quoc}\ \emph {et~al.}(2024)\citenamefont
  {Truong-Quoc}, \citenamefont {Lee}, \citenamefont {Kim},\ and\ \citenamefont
  {Kim}}]{truong2024prediction}%
  \BibitemOpen
  \bibfield  {author} {\bibinfo {author} {\bibfnamefont {C.}~\bibnamefont
  {Truong-Quoc}}, \bibinfo {author} {\bibfnamefont {J.~Y.}\ \bibnamefont
  {Lee}}, \bibinfo {author} {\bibfnamefont {K.~S.}\ \bibnamefont {Kim}},\ and\
  \bibinfo {author} {\bibfnamefont {D.-N.}\ \bibnamefont {Kim}},\ }\bibfield
  {title} {\bibinfo {title} {Prediction of {DNA} origami shape using graph
  neural network},\ }\href@noop {} {\bibfield  {journal} {\bibinfo  {journal}
  {Nature Materials}\ ,\ \bibinfo {pages} {1}} (\bibinfo {year}
  {2024})}\BibitemShut {NoStop}%
\bibitem [{\citenamefont {Karner}\ \emph
  {et~al.}(2020{\natexlab{a}})\citenamefont {Karner}, \citenamefont {Dellago},\
  and\ \citenamefont {Bianchi}}]{karner2020patchiness}%
  \BibitemOpen
  \bibfield  {author} {\bibinfo {author} {\bibfnamefont {C.}~\bibnamefont
  {Karner}}, \bibinfo {author} {\bibfnamefont {C.}~\bibnamefont {Dellago}},\
  and\ \bibinfo {author} {\bibfnamefont {E.}~\bibnamefont {Bianchi}},\
  }\bibfield  {title} {\bibinfo {title} {How patchiness controls the properties
  of chain-like assemblies of colloidal platelets},\ }\href@noop {} {\bibfield
  {journal} {\bibinfo  {journal} {Journal of Physics: Condensed Matter}\
  }\textbf {\bibinfo {volume} {32}},\ \bibinfo {pages} {204001} (\bibinfo
  {year} {2020}{\natexlab{a}})}\BibitemShut {NoStop}%
\bibitem [{\citenamefont {Karner}\ and\ \citenamefont
  {Bianchi}(2024)}]{karner2024anisotropic}%
  \BibitemOpen
  \bibfield  {author} {\bibinfo {author} {\bibfnamefont {C.}~\bibnamefont
  {Karner}}\ and\ \bibinfo {author} {\bibfnamefont {E.}~\bibnamefont
  {Bianchi}},\ }\bibfield  {title} {\bibinfo {title} {Anisotropic
  functionalized platelets: percolation, porosity and network properties},\
  }\href@noop {} {\bibfield  {journal} {\bibinfo  {journal} {Nanoscale
  Advances}\ }\textbf {\bibinfo {volume} {6}},\ \bibinfo {pages} {443}
  (\bibinfo {year} {2024})}\BibitemShut {NoStop}%
\bibitem [{\citenamefont {Xiouras}\ \emph {et~al.}(2018)\citenamefont
  {Xiouras}, \citenamefont {Fytopoulos}, \citenamefont {Jordens}, \citenamefont
  {Boudouvis}, \citenamefont {Van~Gerven},\ and\ \citenamefont
  {Stefanidis}}]{xiouras2018applications}%
  \BibitemOpen
  \bibfield  {author} {\bibinfo {author} {\bibfnamefont {C.}~\bibnamefont
  {Xiouras}}, \bibinfo {author} {\bibfnamefont {A.}~\bibnamefont {Fytopoulos}},
  \bibinfo {author} {\bibfnamefont {J.}~\bibnamefont {Jordens}}, \bibinfo
  {author} {\bibfnamefont {A.~G.}\ \bibnamefont {Boudouvis}}, \bibinfo {author}
  {\bibfnamefont {T.}~\bibnamefont {Van~Gerven}},\ and\ \bibinfo {author}
  {\bibfnamefont {G.~D.}\ \bibnamefont {Stefanidis}},\ }\bibfield  {title}
  {\bibinfo {title} {Applications of ultrasound to chiral crystallization,
  resolution and deracemization},\ }\href@noop {} {\bibfield  {journal}
  {\bibinfo  {journal} {Ultrasonics Sonochemistry}\ }\textbf {\bibinfo {volume}
  {43}},\ \bibinfo {pages} {184} (\bibinfo {year} {2018})}\BibitemShut
  {NoStop}%
\bibitem [{\citenamefont {S{\"o}g{\"u}toglu}\ \emph {et~al.}(2015)\citenamefont
  {S{\"o}g{\"u}toglu}, \citenamefont {Steendam}, \citenamefont {Meekes},
  \citenamefont {Vlieg},\ and\ \citenamefont {Rutjes}}]{sogutoglu2015viedma}%
  \BibitemOpen
  \bibfield  {author} {\bibinfo {author} {\bibfnamefont {L.-C.}\ \bibnamefont
  {S{\"o}g{\"u}toglu}}, \bibinfo {author} {\bibfnamefont {R.~R.}\ \bibnamefont
  {Steendam}}, \bibinfo {author} {\bibfnamefont {H.}~\bibnamefont {Meekes}},
  \bibinfo {author} {\bibfnamefont {E.}~\bibnamefont {Vlieg}},\ and\ \bibinfo
  {author} {\bibfnamefont {F.~P.}\ \bibnamefont {Rutjes}},\ }\bibfield  {title}
  {\bibinfo {title} {Viedma ripening: a reliable crystallisation method to
  reach single chirality},\ }\href@noop {} {\bibfield  {journal} {\bibinfo
  {journal} {Chemical Society Reviews}\ }\textbf {\bibinfo {volume} {44}},\
  \bibinfo {pages} {6723} (\bibinfo {year} {2015})}\BibitemShut {NoStop}%
\bibitem [{\citenamefont {Jacques}\ \emph {et~al.}(1981)\citenamefont
  {Jacques}, \citenamefont {Collet}, \citenamefont {Wilen},\ and\ \citenamefont
  {Collet}}]{jacques1981enantiomers}%
  \BibitemOpen
  \bibfield  {author} {\bibinfo {author} {\bibfnamefont {J.}~\bibnamefont
  {Jacques}}, \bibinfo {author} {\bibfnamefont {A.}~\bibnamefont {Collet}},
  \bibinfo {author} {\bibfnamefont {S.~H.}\ \bibnamefont {Wilen}},\ and\
  \bibinfo {author} {\bibfnamefont {A.}~\bibnamefont {Collet}},\ }\href@noop {}
  {\emph {\bibinfo {title} {Enantiomers, racemates, and resolutions}}}\
  (\bibinfo  {publisher} {Wiley New York},\ \bibinfo {year} {1981})\BibitemShut
  {NoStop}%
\bibitem [{\citenamefont {Lorenz}\ and\ \citenamefont
  {Seidel-Morgenstern}(2014)}]{lorenz2014processes}%
  \BibitemOpen
  \bibfield  {author} {\bibinfo {author} {\bibfnamefont {H.}~\bibnamefont
  {Lorenz}}\ and\ \bibinfo {author} {\bibfnamefont {A.}~\bibnamefont
  {Seidel-Morgenstern}},\ }\bibfield  {title} {\bibinfo {title} {Processes to
  separate enantiomers},\ }\href@noop {} {\bibfield  {journal} {\bibinfo
  {journal} {Angewandte Chemie International Edition}\ }\textbf {\bibinfo
  {volume} {53}},\ \bibinfo {pages} {1218} (\bibinfo {year}
  {2014})}\BibitemShut {NoStop}%
\bibitem [{\citenamefont {Lenz}\ and\ \citenamefont
  {Witten}(2017)}]{lenz2017geometrical}%
  \BibitemOpen
  \bibfield  {author} {\bibinfo {author} {\bibfnamefont {M.}~\bibnamefont
  {Lenz}}\ and\ \bibinfo {author} {\bibfnamefont {T.~A.}\ \bibnamefont
  {Witten}},\ }\bibfield  {title} {\bibinfo {title} {Geometrical frustration
  yields fibre formation in self-assembly},\ }\href@noop {} {\bibfield
  {journal} {\bibinfo  {journal} {Nature physics}\ }\textbf {\bibinfo {volume}
  {13}},\ \bibinfo {pages} {1100} (\bibinfo {year} {2017})}\BibitemShut
  {NoStop}%
\bibitem [{\citenamefont {Serafin}\ \emph {et~al.}(2021)\citenamefont
  {Serafin}, \citenamefont {Lu}, \citenamefont {Kotov}, \citenamefont {Sun},\
  and\ \citenamefont {Mao}}]{serafin2021frustrated}%
  \BibitemOpen
  \bibfield  {author} {\bibinfo {author} {\bibfnamefont {F.}~\bibnamefont
  {Serafin}}, \bibinfo {author} {\bibfnamefont {J.}~\bibnamefont {Lu}},
  \bibinfo {author} {\bibfnamefont {N.}~\bibnamefont {Kotov}}, \bibinfo
  {author} {\bibfnamefont {K.}~\bibnamefont {Sun}},\ and\ \bibinfo {author}
  {\bibfnamefont {X.}~\bibnamefont {Mao}},\ }\bibfield  {title} {\bibinfo
  {title} {Frustrated self-assembly of non-euclidean crystals of
  nanoparticles},\ }\href@noop {} {\bibfield  {journal} {\bibinfo  {journal}
  {Nature Communications}\ }\textbf {\bibinfo {volume} {12}},\ \bibinfo {pages}
  {4925} (\bibinfo {year} {2021})}\BibitemShut {NoStop}%
\bibitem [{\citenamefont {Tyukodi}\ \emph {et~al.}(2022)\citenamefont
  {Tyukodi}, \citenamefont {Mohajerani}, \citenamefont {Hall}, \citenamefont
  {Grason},\ and\ \citenamefont {Hagan}}]{tyukodi2022thermodynamic}%
  \BibitemOpen
  \bibfield  {author} {\bibinfo {author} {\bibfnamefont {B.}~\bibnamefont
  {Tyukodi}}, \bibinfo {author} {\bibfnamefont {F.}~\bibnamefont {Mohajerani}},
  \bibinfo {author} {\bibfnamefont {D.~M.}\ \bibnamefont {Hall}}, \bibinfo
  {author} {\bibfnamefont {G.~M.}\ \bibnamefont {Grason}},\ and\ \bibinfo
  {author} {\bibfnamefont {M.~F.}\ \bibnamefont {Hagan}},\ }\bibfield  {title}
  {\bibinfo {title} {Thermodynamic size control in curvature-frustrated
  tubules: Self-limitation with open boundaries},\ }\href@noop {} {\bibfield
  {journal} {\bibinfo  {journal} {ACS nano}\ }\textbf {\bibinfo {volume}
  {16}},\ \bibinfo {pages} {9077} (\bibinfo {year} {2022})}\BibitemShut
  {NoStop}%
\bibitem [{\citenamefont {Carpenter}\ and\ \citenamefont
  {Gr\"unwald}(2020)}]{carpenter2020heterogeneous}%
  \BibitemOpen
  \bibfield  {author} {\bibinfo {author} {\bibfnamefont {J.~E.}\ \bibnamefont
  {Carpenter}}\ and\ \bibinfo {author} {\bibfnamefont {M.}~\bibnamefont
  {Gr\"unwald}},\ }\bibfield  {title} {\bibinfo {title} {Heterogeneous
  interactions promote crystallization and spontaneous resolution of chiral
  molecules},\ }\href@noop {} {\bibfield  {journal} {\bibinfo  {journal}
  {Journal of the American Chemical Society}\ }\textbf {\bibinfo {volume}
  {142}},\ \bibinfo {pages} {10755} (\bibinfo {year} {2020})}\BibitemShut
  {NoStop}%
\bibitem [{\citenamefont {Blunt}\ \emph {et~al.}(2008)\citenamefont {Blunt},
  \citenamefont {Russell}, \citenamefont {del Carmen Gim{\'e}nez-L{\'o}pez},
  \citenamefont {Garrahan}, \citenamefont {Lin}, \citenamefont {Schr{\"o}der},
  \citenamefont {Champness},\ and\ \citenamefont {Beton}}]{Science_2008}%
  \BibitemOpen
  \bibfield  {author} {\bibinfo {author} {\bibfnamefont {M.~O.}\ \bibnamefont
  {Blunt}}, \bibinfo {author} {\bibfnamefont {J.~C.}\ \bibnamefont {Russell}},
  \bibinfo {author} {\bibfnamefont {M.}~\bibnamefont {del Carmen
  Gim{\'e}nez-L{\'o}pez}}, \bibinfo {author} {\bibfnamefont {J.~P.}\
  \bibnamefont {Garrahan}}, \bibinfo {author} {\bibfnamefont {X.}~\bibnamefont
  {Lin}}, \bibinfo {author} {\bibfnamefont {M.}~\bibnamefont {Schr{\"o}der}},
  \bibinfo {author} {\bibfnamefont {N.~R.}\ \bibnamefont {Champness}},\ and\
  \bibinfo {author} {\bibfnamefont {P.~H.}\ \bibnamefont {Beton}},\ }\bibfield
  {title} {\bibinfo {title} {Random tiling and topological defects in a
  two-dimensional molecular network},\ }\href@noop {} {\bibfield  {journal}
  {\bibinfo  {journal} {Science}\ }\textbf {\bibinfo {volume} {322}},\ \bibinfo
  {pages} {1077} (\bibinfo {year} {2008})}\BibitemShut {NoStop}%
\bibitem [{\citenamefont {Bernstein}(2020)}]{bernstein2020polymorphism}%
  \BibitemOpen
  \bibfield  {author} {\bibinfo {author} {\bibfnamefont {J.}~\bibnamefont
  {Bernstein}},\ }\href@noop {} {\emph {\bibinfo {title} {Polymorphism in
  Molecular Crystals 2e}}},\ Vol.~\bibinfo {volume} {30}\ (\bibinfo
  {publisher} {International Union of Crystal},\ \bibinfo {year}
  {2020})\BibitemShut {NoStop}%
\bibitem [{\citenamefont {Cruz-Cabeza}\ \emph {et~al.}(2015)\citenamefont
  {Cruz-Cabeza}, \citenamefont {Reutzel-Edens},\ and\ \citenamefont
  {Bernstein}}]{cruz2015facts}%
  \BibitemOpen
  \bibfield  {author} {\bibinfo {author} {\bibfnamefont {A.~J.}\ \bibnamefont
  {Cruz-Cabeza}}, \bibinfo {author} {\bibfnamefont {S.~M.}\ \bibnamefont
  {Reutzel-Edens}},\ and\ \bibinfo {author} {\bibfnamefont {J.}~\bibnamefont
  {Bernstein}},\ }\bibfield  {title} {\bibinfo {title} {Facts and fictions
  about polymorphism},\ }\href@noop {} {\bibfield  {journal} {\bibinfo
  {journal} {Chemical Society Reviews}\ }\textbf {\bibinfo {volume} {44}},\
  \bibinfo {pages} {8619} (\bibinfo {year} {2015})}\BibitemShut {NoStop}%
\bibitem [{\citenamefont {Lee}\ \emph {et~al.}(2011)\citenamefont {Lee},
  \citenamefont {Erdemir},\ and\ \citenamefont {Myerson}}]{lee2011crystal}%
  \BibitemOpen
  \bibfield  {author} {\bibinfo {author} {\bibfnamefont {A.~Y.}\ \bibnamefont
  {Lee}}, \bibinfo {author} {\bibfnamefont {D.}~\bibnamefont {Erdemir}},\ and\
  \bibinfo {author} {\bibfnamefont {A.~S.}\ \bibnamefont {Myerson}},\
  }\bibfield  {title} {\bibinfo {title} {Crystal polymorphism in chemical
  process development},\ }\href@noop {} {\bibfield  {journal} {\bibinfo
  {journal} {Annual review of chemical and biomolecular engineering}\ }\textbf
  {\bibinfo {volume} {2}},\ \bibinfo {pages} {259} (\bibinfo {year}
  {2011})}\BibitemShut {NoStop}%
\bibitem [{\citenamefont {Calcaterra}\ and\ \citenamefont
  {D’Acquarica}(2018)}]{calcaterra2018market}%
  \BibitemOpen
  \bibfield  {author} {\bibinfo {author} {\bibfnamefont {A.}~\bibnamefont
  {Calcaterra}}\ and\ \bibinfo {author} {\bibfnamefont {I.}~\bibnamefont
  {D’Acquarica}},\ }\bibfield  {title} {\bibinfo {title} {The market of
  chiral drugs: Chiral switches versus de novo enantiomerically pure
  compounds},\ }\href@noop {} {\bibfield  {journal} {\bibinfo  {journal}
  {Journal of pharmaceutical and biomedical analysis}\ }\textbf {\bibinfo
  {volume} {147}},\ \bibinfo {pages} {323} (\bibinfo {year}
  {2018})}\BibitemShut {NoStop}%
\bibitem [{\citenamefont {Nguyen}\ \emph {et~al.}(2006)\citenamefont {Nguyen},
  \citenamefont {He},\ and\ \citenamefont {Pham-Huy}}]{nguyen2006chiral}%
  \BibitemOpen
  \bibfield  {author} {\bibinfo {author} {\bibfnamefont {L.~A.}\ \bibnamefont
  {Nguyen}}, \bibinfo {author} {\bibfnamefont {H.}~\bibnamefont {He}},\ and\
  \bibinfo {author} {\bibfnamefont {C.}~\bibnamefont {Pham-Huy}},\ }\bibfield
  {title} {\bibinfo {title} {Chiral drugs: an overview},\ }\href@noop {}
  {\bibfield  {journal} {\bibinfo  {journal} {International journal of
  biomedical science: IJBS}\ }\textbf {\bibinfo {volume} {2}},\ \bibinfo
  {pages} {85} (\bibinfo {year} {2006})}\BibitemShut {NoStop}%
\bibitem [{\citenamefont {Smith}(2009)}]{smith2009chiral}%
  \BibitemOpen
  \bibfield  {author} {\bibinfo {author} {\bibfnamefont {S.~W.}\ \bibnamefont
  {Smith}},\ }\bibfield  {title} {\bibinfo {title} {Chiral toxicology: it's the
  same thing… only different},\ }\href@noop {} {\bibfield  {journal}
  {\bibinfo  {journal} {Toxicological sciences}\ }\textbf {\bibinfo {volume}
  {110}},\ \bibinfo {pages} {4} (\bibinfo {year} {2009})}\BibitemShut {NoStop}%
\bibitem [{\citenamefont {Golshtein}\ and\ \citenamefont
  {Tretyakov}(1996)}]{Golshtein_1996}%
  \BibitemOpen
  \bibfield  {author} {\bibinfo {author} {\bibfnamefont {E.~G.}\ \bibnamefont
  {Golshtein}}\ and\ \bibinfo {author} {\bibfnamefont {N.}~\bibnamefont
  {Tretyakov}},\ }\href@noop {} {\emph {\bibinfo {title} {Modified Lagrangians
  and monotone maps in optimization}}}\ (\bibinfo  {publisher} {Wiley},\
  \bibinfo {year} {1996})\BibitemShut {NoStop}%
\bibitem [{\citenamefont {Karner}\ \emph
  {et~al.}(2020{\natexlab{b}})\citenamefont {Karner}, \citenamefont
  {M{\"u}ller},\ and\ \citenamefont {Bianchi}}]{karner2020matter}%
  \BibitemOpen
  \bibfield  {author} {\bibinfo {author} {\bibfnamefont {C.}~\bibnamefont
  {Karner}}, \bibinfo {author} {\bibfnamefont {F.}~\bibnamefont {M{\"u}ller}},\
  and\ \bibinfo {author} {\bibfnamefont {E.}~\bibnamefont {Bianchi}},\
  }\bibfield  {title} {\bibinfo {title} {A matter of size and placement:
  Varying the patch size of anisotropic patchy colloids},\ }\href@noop {}
  {\bibfield  {journal} {\bibinfo  {journal} {International Journal of
  Molecular Sciences}\ }\textbf {\bibinfo {volume} {21}},\ \bibinfo {pages}
  {8621} (\bibinfo {year} {2020}{\natexlab{b}})}\BibitemShut {NoStop}%
\bibitem [{\citenamefont {Whitelam}\ and\ \citenamefont
  {Geissler}(2007)}]{Whitelam2007}%
  \BibitemOpen
  \bibfield  {author} {\bibinfo {author} {\bibfnamefont {S.}~\bibnamefont
  {Whitelam}}\ and\ \bibinfo {author} {\bibfnamefont {P.}~\bibnamefont
  {Geissler}},\ }\bibfield  {title} {\bibinfo {title} {Avoiding unphysical
  kinetic traps in {Monte Carlo} simulations of strongly attractive
  particles},\ }\href@noop {} {\bibfield  {journal} {\bibinfo  {journal}
  {Journal of Chemical Physics}\ }\textbf {\bibinfo {volume} {127}},\ \bibinfo
  {pages} {154101} (\bibinfo {year} {2007})}\BibitemShut {NoStop}%
\bibitem [{\citenamefont {Whitelam}(2011)}]{Whitelam2010}%
  \BibitemOpen
  \bibfield  {author} {\bibinfo {author} {\bibfnamefont {S.}~\bibnamefont
  {Whitelam}},\ }\bibfield  {title} {\bibinfo {title} {Approximating the
  dynamical evolution of systems of strongly interacting overdamped
  particles},\ }\href@noop {} {\bibfield  {journal} {\bibinfo  {journal}
  {Molecular Simulation}\ }\textbf {\bibinfo {volume} {37}},\ \bibinfo {pages}
  {606} (\bibinfo {year} {2011})}\BibitemShut {NoStop}%
\bibitem [{\citenamefont {Steinhardt}\ \emph {et~al.}(1983)\citenamefont
  {Steinhardt}, \citenamefont {Nelson},\ and\ \citenamefont
  {Ronchetti}}]{steinhardt1983bond}%
  \BibitemOpen
  \bibfield  {author} {\bibinfo {author} {\bibfnamefont {P.~J.}\ \bibnamefont
  {Steinhardt}}, \bibinfo {author} {\bibfnamefont {D.~R.}\ \bibnamefont
  {Nelson}},\ and\ \bibinfo {author} {\bibfnamefont {M.}~\bibnamefont
  {Ronchetti}},\ }\bibfield  {title} {\bibinfo {title} {Bond-orientational
  order in liquids and glasses},\ }\href@noop {} {\bibfield  {journal}
  {\bibinfo  {journal} {Physical Review B}\ }\textbf {\bibinfo {volume} {28}},\
  \bibinfo {pages} {784} (\bibinfo {year} {1983})}\BibitemShut {NoStop}%
\end{thebibliography}

\end{document}